\documentclass[aps,prd,twocolumn,preprintnumbers,superscriptaddress,nofootinbib,floatfix]{revtex4-1}

\usepackage[dvipsnames]{xcolor}
\usepackage{amssymb,amsmath,amsfonts}
\usepackage{bm}
\usepackage{graphicx}
\usepackage{epstopdf}
\usepackage{hyperref}
\usepackage{array}
\usepackage{colortbl}
\usepackage[utf8]{inputenc}
\usepackage{soul}
\usepackage{color}
\usepackage{booktabs}
\usepackage[T1]{fontenc}
\usepackage{longtable}
\usepackage[normalem]{ulem}

\enlargethispage{\baselineskip}

\definecolor{steelblue}{RGB}{25,25,112}
\definecolor{dullblue}{rgb}{0,0.298,0.49}
\definecolor{darkred}{rgb}{0.545,0,0}
\definecolor{blue2}{cmyk}{1, 0.1, 0.1, 0}

\hypersetup{colorlinks,linkcolor={darkred},citecolor={dullblue},urlcolor={dullblue}} 

\everymath{\displaystyle} 
\renewcommand\({\left(}
\renewcommand\){\right)}
\renewcommand\[{\left[}
\renewcommand\]{\right]}
\newcommand{\be}{\begin{equation}}
\newcommand{\ee}{\end{equation}}
\newcommand{\bea}{\begin{eqnarray}}
\newcommand{\eea}{\end{eqnarray}}
\newcommand{\DD}{\mathcal D}
\newcommand{\dbd}[2]{\frac{\mathrm{d}#1}{\mathrm{d}#2}}
\newcommand{\LL}{\mathcal L}
\newcommand{\MW}{^\mathrm{MW}}
\newcommand{\SIDDMW}{\sigma_p^{\rm SI; MW}}
\newcommand{\disk}{^\mathrm{disk}}
\newcommand{\SIDDdisk}{\sigma_p^{\rm SI; disk}}
\newcommand{\SH}{^\mathrm{sh}}
\newcommand{\SIDDSH}{\sigma_p^{\rm SI; sh}}

\newcommand{\dd}{\mathrm{d}}

\usepackage{array}
\newcolumntype{L}[1]{>{\raggedright\let\newline\\\arraybackslash\hspace{0pt}}m{#1}}
\newcolumntype{C}[1]{>{\centering\let\newline\\\arraybackslash\hspace{0pt}}m{#1}}
\newcolumntype{R}[1]{>{\raggedleft\let\newline\\\arraybackslash\hspace{0pt}}m{#1}}

\newcommand{\linkPaSpec}{https://github.com/sbaum90/paleoSpec}
\newcommand{\linkPaSens}{https://github.com/sbaum90/paleoSens}

\begin{document}

\title{Galactic Geology:\\
Probing Time-Varying Dark Matter Signals with Paleo-Detectors}

\newcommand{\OKC}{\affiliation{The Oskar Klein Centre, Department of Physics, Stockholm University, AlbaNova, SE-106 91 Stockholm, Sweden}} 
\newcommand{\Stanford}{\affiliation{Stanford Institute for Theoretical Physics, Stanford University, Stanford, CA 94305, USA}} 
\author{Sebastian Baum}\email[Electronic address: ]{sbaum@stanford.edu} \Stanford
\author{William DeRocco}\email[Electronic address: ]{wderocco@stanford.edu} \Stanford
\author{Thomas D. P. Edwards}\email[Electronic address: ]{thomas.edwards@fysik.su.se} \OKC
\author{Saarik Kalia}\email[Electronic address: ]{saarik@stanford.edu} \Stanford

\date{\today}

\begin{abstract}
Paleo-detectors are a proposed experimental technique to search for dark matter by reading out the damage tracks caused by nuclear recoils in small samples of natural minerals. Unlike a conventional real-time direct detection experiment, paleo-detectors have been accumulating these tracks for up to a billion years. These long integration times offer a unique possibility: by reading out paleo-detectors of different ages, one can explore the {\it time-variation} of signals on megayear to gigayear timescales. We investigate two examples of dark matter substructure that could give rise to such time-varying signals. First, a dark disk through which the Earth would pass every $\sim$45\,Myr, and second, a dark matter subhalo that the Earth encountered during the past gigayear. We demonstrate that paleo-detectors are sensitive to these examples under a wide variety of experimental scenarios, even in the presence of substantial background uncertainties. This paper shows that paleo-detectors may hold the key to unraveling our Galactic history.
\end{abstract}

\maketitle

\section{Introduction} \label{sec:intro}

Many naturally-occurring minerals are excellent nuclear recoil detectors. When an atomic nucleus within the mineral receives a ``kick'', it travels through the crystal and leaves a persistent \textit{damage track}~\cite{Fleischer:1964,Fleischer383,Fleischer:1965yv,GUO2012233}. Minerals found on Earth are up to $\mathcal{O}(1)\,$Gyr old and have been recording damage tracks over their entire age.\footnote{We use ``age'' to describe the time over which a mineral has been recording nuclear damage tracks, which can be different to the time since formation. For example, for a sample that has recrystallized, ``age'' refers to the time to the last recrystallization.} The idea of leveraging the long exposure times of natural minerals to explore rare events has long been explored in the literature~\cite{Goto:1958,Goto:1963zz,Fleischer:1969mj,Fleischer:1970zy,Fleischer:1970vm,Alvarez:1970zu,Kolm:1971xb,Eberhard:1971re,Ross:1973it,Price:1983ax,Kovalik:1986zz,Price:1986ky,Ghosh:1990ki,Jeon:1995rf,SnowdenIfft:1995ke,Collar:1994mj,Engel:1995gw,SnowdenIfft:1997hd,Collar:1999md}. However, modern microscopy techniques promise damage track readout resolutions of $\mathcal{O}(1-10)\,$nm in samples as large as $\mathcal{O}(0.01-100)\,$g. The idea of using such modern microscopy techniques to search for dark matter (DM) or neutrino induced recoil tracks in natural minerals has been dubbed \textit{paleo-detectors}~\cite{Baum:2018tfw, Drukier:2018pdy, Edwards:2018hcf, Baum:2019fqm, Jordan:2020gxx, Arellano:2021jul,Baum:2021jak} (see also Refs.~\cite{Essig:2016crl,Budnik:2017sbu,Rajendran:2017ynw,Sidhu:2019qoa,Lehmann:2019zgt,Bhoonah:2020fys,Cogswell:2021qlq,Ebadi:2021cte,Acevedo:2021tbl} for related recent work). For example, hard X-ray microscopy~\cite{RODRIGUEZ2014150,SAXS3d,SAXSres} could allow for the readout of $\mathcal{O}(100)\,$g of material with track-length resolution of $\mathcal{O}(10)\,$nm. Such resolution corresponds to a nuclear recoil energy threshold of $\mathcal{O}(1)\,$keV, comparable to the threshold of liquid-Xe-based direct detection experiments~\cite{Schumann:2019eaa}. Reading out $\mathcal{O}(100)\,$g of a $1\,$Gyr old sample would lead to an exposure of $\varepsilon = 100\,{\rm g\,Gyr} = 10^4\,\mathrm{tonne}\times \mathrm{year}$, orders of magnitude larger than the $\mathcal{O}(1-10)\,\mathrm{tonne}\times \mathrm{year}$ exposures of conventional direct detection experiments~\cite{Schumann:2019eaa,Angloher:2015ewa, Aprile:2015uzo, Armengaud:2016cvl, Aalbers:2016jon, Akerib:2016vxi, Mount:2017qzi, Agnese:2017jvy, Aalseth:2017fik, Petricca:2017zdp, Amaudruz:2017ibl, Agnes:2018fwg, Aprile:2018dbl, Armengaud:2019kfj, Wang:2020coa}. This combination of low threshold and large exposure provides a unique opportunity to explore physics which gives rise to rare nuclear recoils such as DM~\cite{Baum:2018tfw, Drukier:2018pdy, Edwards:2018hcf,Baum:2021jak} and neutrinos produced in the Sun~\cite{Arellano:2021jul}, in Galactic supernovae~\cite{Baum:2019fqm}, and by cosmic rays interacting with Earth's atmosphere~\cite{Jordan:2020gxx}.

\begin{figure}
    \includegraphics[width=\linewidth]{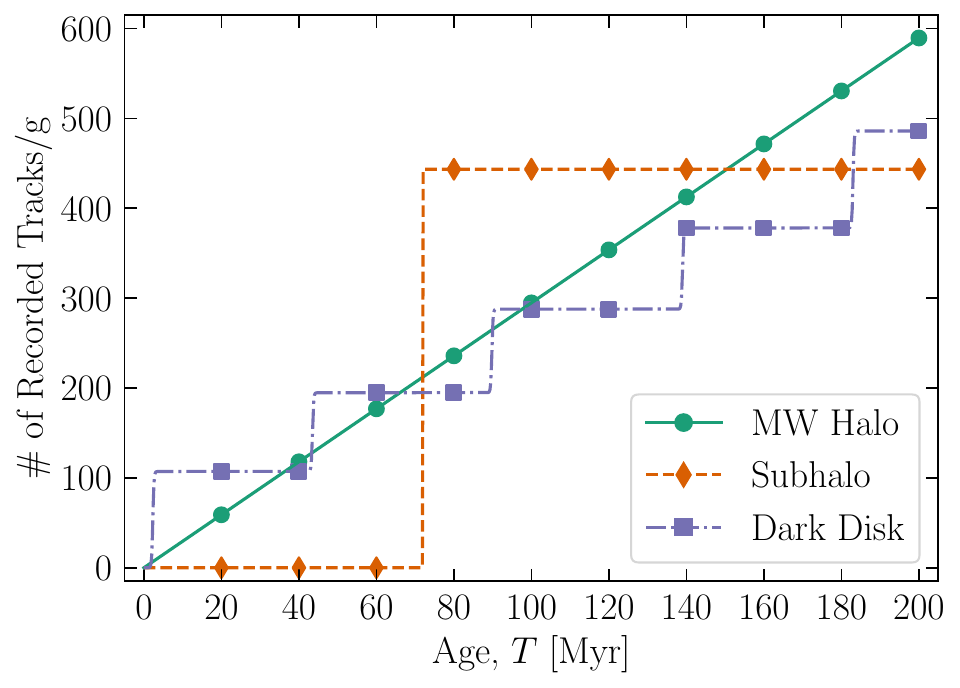}
    \caption{Illustration of the time-dependence of the number of damage tracks which would be recorded in a gram-sized paleo-detector of age $T$ for three different DM signals: the smooth Milky Way halo (solid green), a DM subhalo Earth traversed $T\SH \sim 70\,$Myr ago (dashed orange), and the periodic crossings through a dark disk (dash-dotted purple). The markers indicate a possible series of samples of different ages. We choose illustrative values for the various signal parameters (described in Sec.~\ref{sec:signals}) in order to obtain comparable numbers of events. Note that a paleo-detector would actually record the sum of the Milky Way halo and either subhalo or dark disk contributions, along with contributions from various backgrounds (see Sec.~\ref{sec:basics}).}
    \label{Fig:timevar}
\end{figure}

The long exposure times of paleo-detectors offer an additional unique feature; by using a series of paleo-detectors with different ages, one can probe the temporal dependence of signals that evolve over Myr to Gyr timescales. This is because any sample will contain the integrated number of tracks recorded over its age. Previous work~\cite{Baum:2019fqm, Jordan:2020gxx, Arellano:2021jul} has taken first steps to exploring the sensitivity of paleo-detectors to time-varying signals, but has not developed a robust framework to quantitatively study the sensitivity to such signals in the presence of experimental and modeling uncertainties. Here, we develop a general framework to explore this sensitivity and demonstrate it on two examples arising from DM substructure, illustrated in Fig.~\ref{Fig:timevar}: 
\begin{itemize}
    \item Periodic transits through a \textit{dark disk}, 
    \item A past transit through a DM \textit{subhalo}.
\end{itemize}
Although the main aim of this paper is to demonstrate the sensitivity of paleo-detectors to time-varying signals, the two substructure scenarios we consider are of direct interest to the DM community. 

If a component of DM is able to dissipate energy (see Refs.~\cite{Agrawal:2017rvu, Rosenberg:2017qia, Foot:2016wvj, Buckley:2017ttd, Cline:2013pca, Boddy:2016bbu, Schutz:2014nka, CyrRacine:2012fz,Chacko:2021vin} for examples), a thin dark disk co-planar with the Galactic baryonic disk could form~\cite{Fan:2013yva}. The Solar System oscillates normal to the disk plane with a period of $\sim 90\,$Myr, with the last mid-plane crossing happening $\sim 2.3\,\mathrm{Myr}$ ago. Thus, paleo-detectors would see a series of injections of tracks as illustrated in Fig.~\ref{Fig:timevar}. Astrometric measurements of stars are sensitive to the gravitational effects of a dark disk and provide upper limits on its surface density, $\Sigma\disk \lesssim 5\,M_\odot/{\rm pc}^2$~\cite{Kramer:2016dqu,Schutz:2017tfp,Widmark:2018ylf,Buch:2018qdr,Widmark:2021gqx}, although they are subject to a host of uncertainties~\cite{Kramer:2016dew,Widmark:2020vqi}. We will show that, depending on the scattering cross section of the DM making up the dark disk, one could probe dramatically lower surface densities using a series of paleo-detectors of different ages. 

In contrast to a dark disk, subhalos are a generic expectation of cold DM in standard cosmologies. The growth of DM halos is described by hierarchical structure formation~\cite{Jiang:2014nsa, vandenBosch:2004zs, Giocoli:2007gf, Gao:2004au, Dolag:2008ar, Springel:2008cc, 1974ApJ...187..425P} which results in a power-law halo mass function (the number of halos, $N$, per halo mass, $M$), $\dd N/\dd M\propto M^{-\alpha}$, with $\alpha \sim 1.8-2$~\cite{Hiroshima:2018kfv}. Any isolated {\it field-halo} contains a population of {\it subhalos}, whose mass function, spatial distribution, and density profiles are influenced by the tidal force of their host galaxy (see, for example, Refs.~\cite{Sanchez-Conde:2013yxa,Moline:2016pbm,Hiroshima:2018kfv,Ando:2019xlm,Wang:2019ftp}). Astronomical observations constrain the halo mass function down to scales of the order $M \sim 10^7\,M_\odot$ (see, for example, Refs.~\cite{Nadler:2019zrb,Schutz:2020jox,Nadler:2020prv,Mao:2020rga,Das:2020nwc,Maamari:2020aqz,Nadler:2021dft} for recent work); however, at smaller masses, $\dd N/\dd M$ is essentially unconstrained. The subhalo mass function at these small scales contains crucial information about both the DM model and early Universe cosmology~\cite{Stafford:2020ppr,Blinov:2021axd}. By using a series of paleo-detectors of different ages, one could be sensitive to transits through subhalos over the last Gyr. While we find the chance of detecting a subhalo encounter with paleo-detectors to be rather low  (see Appendix~\ref{app:subhalo}) assuming a mass function arising from standard cosmology~\cite{Moline:2016pbm}, the mass function can be significantly enhanced by nonstandard cosmologies~\cite{Sanati2020,Halpern2015}. Thus the detection of a subhalo transit could not only probe the subhalo mass function in an unconstrained mass range, but also open a new window to the cosmology of the early Universe.

Crucially, these two examples would lead to very different time-dependence of the signals. A dark disk would induce damage tracks periodically every $\sim 45\,$Myr, while a single subhalo encounter leads to all associated tracks being recorded practically at once, see Fig.~\ref{Fig:timevar}. The temporal dependence of either of these signals is distinct from the MW halo, which would induce tracks at a constant rate. While we focus on these two particular examples, the results are general --- paleo-detectors offer a unique and powerful tool to explore time-varying signals. As we will see, paleo-detectors remain sensitive to such time-variations for a wide variety of experimental scenarios and in the presence of modeling uncertainties.

The remainder of this paper is organized as follows: in Sec.~\ref{sec:basics} we discuss the basics of paleo-detectors, including backgrounds and the calculation of track length spectra. Section~\ref{sec:signals} discusses the signal model for both the dark disk and subhalo encounters. In Sec.~\ref{sec:stat}, we describe the statistical procedure used to estimate the sensitivity of a series of paleo-detectors to time-varying signals. In Sec.~\ref{sec:results}, we show sensitivity projections for the {\it dark disk} (Sec.~\ref{sec:DarkDiskresults}) and {\it subhalo} (Sec.~\ref{sec:Subhaloresults}) scenarios discussed above. In Sec.~\ref{sec:uncertainty}, we estimate the effect of modeling uncertainties on the sensitivity. We conclude in Sec.~\ref{sec:conclusion}. Finally, in Appendix~\ref{app:subhalo}, we discuss the probability of a detectable subhalo encounter, while in Appendix~\ref{app:notation_table} we provide a table detailing our notation throughout the paper. We make the code used in this work available: \href{\linkPaSpec}{\texttt{paleoSpec}}~\cite{PaleoSpec} for the computation of the signal and background spectra, and \href{\linkPaSens}{\texttt{paleoSens}}~\cite{PaleoSens} for the sensitivity forecasts.

\section{Paleo-Detector Basics} \label{sec:basics}

The experimental observable in a paleo-detector is the track length spectrum. In this section, we discuss the basic formalism for computing track length spectra, the two primary readout scenarios we consider in our analyses, the expected background contributions, and some aspects of mineral selection. These issues have been extensively discussed in a series of previous papers~\cite{Baum:2018tfw, Drukier:2018pdy, Edwards:2018hcf, Baum:2019fqm}, and we will describe only the most important aspects here.

\medskip

\textbf{\textit{Track Lengths ---}} A recoiling nucleus leaves a permanent damage track in a solid state nuclear track detector~\cite{Seitz:1949,Fleischer:1964,Fleischer:1965,Fleischer383,Fleischer:1965yv,GUO2012233}. As a proxy for the length of the damage track, $x_T$, for a given nucleus with recoil energy $E_R$, we will use its range,
\begin{equation}\label{eq:tracklength}
    x_T(E_R) = \int_0^{E_R} \left|\frac{\mathrm{d} E}{\mathrm{d} x_T}\right|^{-1} \dd E \;,
\end{equation}
where $\dd E/\dd x_T$ is the stopping power of the nucleus in the target material. The actual length of the damage track may differ from the range if, for example, the nucleus' trajectory is not a straight line or if a lasting damage track is created only along some portion of the length it travels through the material. Previous (numerical) studies suggest that such effects are small~\cite{Drukier:2018pdy}. Furthermore, the effects of thermal annealing could potentially be significant over geological timescales; fortunately, any associated modifications to the track lengths would be similar for both the signal and background recoils. We use the software package \texttt{SRIM}~\cite{Ziegler:1985,Ziegler:2010} to compute the stopping powers; note that analytic estimates of track lengths agree well with the results from \texttt{SRIM}~\cite{Drukier:2018pdy}. 

For any source of nuclear recoils, one typically computes the differential event rate $(\dd R/\dd E_R)_i$ per unit target mass, with respect to recoil energy $E_R$. The rate for each species of constituent nuclei in the target material is indexed by $i$. The differential rate with respect to track length $x_T$ is then obtained by summing over the different isotopes with mass fraction $\xi_i$ and weighting by the associated stopping power,
\begin{equation} \label{eq:dRdx}
    \frac{\mathrm{d}R}{\mathrm{d}x_T} = \sum_i \xi_i \left(\frac{\mathrm{d}R}{\mathrm{d}E_R}\right)_i \left(\frac{\mathrm{d}E_R}{\mathrm{d}x_T}\right)_i\;.
\end{equation}
Throughout this work, we will only include nuclei with mass number $A > 4$ in the sum in Eq.~\eqref{eq:dRdx}. Lighter nuclei (i.e., H and He) do not give rise to permanent damage tracks in typical minerals, see the discussion in~\cite{Baum:2018tfw, Drukier:2018pdy}.

\medskip

\textbf{\textit{Readout ---}} Nuclear damage tracks can be read out using a variety of microscopy techniques, see Ref.~\cite{Drukier:2018pdy} for a discussion. For definiteness, we will consider two scenarios:
\begin{itemize}
    \item {\it High-resolution scenario:} We assume that tracks can be read out with spatial resolution $\sigma_{x_T} = 1\,\mathrm{nm}$ which is potentially achievable with helium-ion beam microscopy~\cite{Hill:2012}. Using focused-ion-beams~\cite{Lombardo:pp5019,Joens:2013} and/or pulsed lasers~\cite{ECHLIN20151,PFEIFENBERGER2017109,Randolph:2018} to remove layers of material which have already been imaged, it should be possible to read out $M_s = 0.01 \,\mathrm{g}$ of material.
    \item {\it Low-resolution scenario}: Using small angle X-ray scattering tomography, track length resolutions of $\sigma_{x_T} = 15\,$nm seem feasible. Fortunately, readout is significantly faster than with helium-ion beam microscopy~\cite{Rodriguez:2014,Holler:2014,Schaff:2015}, meaning that we can consider significantly larger samples, $M_s = 100 \,\mathrm{g}$.
\end{itemize}
The optimal choice of readout method will depend on the signal of interest --- we will discuss our specific choices in Secs.~\ref{sec:stat}-\ref{sec:results}. 

The finite resolution of the track readout process causes the true track length spectra to be \textit{smeared}. We model the rate at which tracks are produced with observed track length $x_T \in \left[ x_i^{\rm min} ,x_i^{\rm max} \right]$ as
\be
\label{eqn:smearing}
    R_i(x_i^{\rm min} ,x_i^{\rm max})=\int_0^\infty W(x_T';x_i^{\rm min} ,x_i^{\rm max} )\frac{\mathrm{d}R}{\mathrm{d}x'_T} \,\dd x_T' \;,
\ee
where $W$ is a window function which describes the smearing. We will assume that the probability of observing a track length $x_T$ for a track with true length $x'_T$ is Gaussian-distributed with variance $\sigma_{x_T}^2$. The corresponding window function is
\be\label{eqn:window}
    W=\frac12 \[ \mathrm{erf} \(\frac{x_T'-x_i^{\rm min}}{\sqrt2\sigma_{x_T}}\) - \mathrm{erf}\(\frac{x_T'-x_i^{\rm max}}{\sqrt2\sigma_{x_T}}\) \]\;.
\ee
Our assumption of the smearing function being well-described by a Gaussian over all track lengths can lead to the problematic case of the unsmeared track length spectra containing \textit{no} tracks above the readout resolution while the smeared track length spectra does. In reality, the smearing function must be calibrated on data and the smallest measurable track length should be investigated. For now, we take a conservative approach and truncate the unsmeared track length spectra, $\dd R/\dd x_T$, at $\sigma_{x_T}/2$ to avoid this problematic case. 

In the remainder of this work, we will use ${\bm R} = \{R_1, \ldots, R_N\}$ to denote the binned and smeared (with respect to track length) recoil rate per unit target mass for bins $i=1,\dots,N$. The observable in a paleo-detector is ultimately the number of tracks in a given bin, $N_i$. To compute $N_i$ from $R_i$, we must integrate over the time the sample has been recording tracks, and multiply with the sample mass, $M_s$,
\begin{equation} \label{eq:Ni}
    N_i = M_s \int_0^T R_i\,\dd t = M_s n_i \;,
\end{equation}
where we have introduced $n_i = {\textstyle \int}_0^T R_i\,\dd t$, the number of tracks per unit target mass in the $i$-th bin. Analogous to ${\bm R}$, we will denote ${\bm n} = \{n_1, \ldots, n_N\}$ and ${\bm N} = \{N_1, \ldots, N_N\}$. Note that one can exchange the order of the integrals and the summation in Eqs.~\eqref{eq:dRdx}--\eqref{eq:Ni} and calculate ${\bm n}$ from $(\dd n/\dd E_R)_i = {\textstyle \int}_0^T (\dd R/\dd E_R)_i\,\dd t$.

\begin{figure*}
    \includegraphics[width=0.49\linewidth]{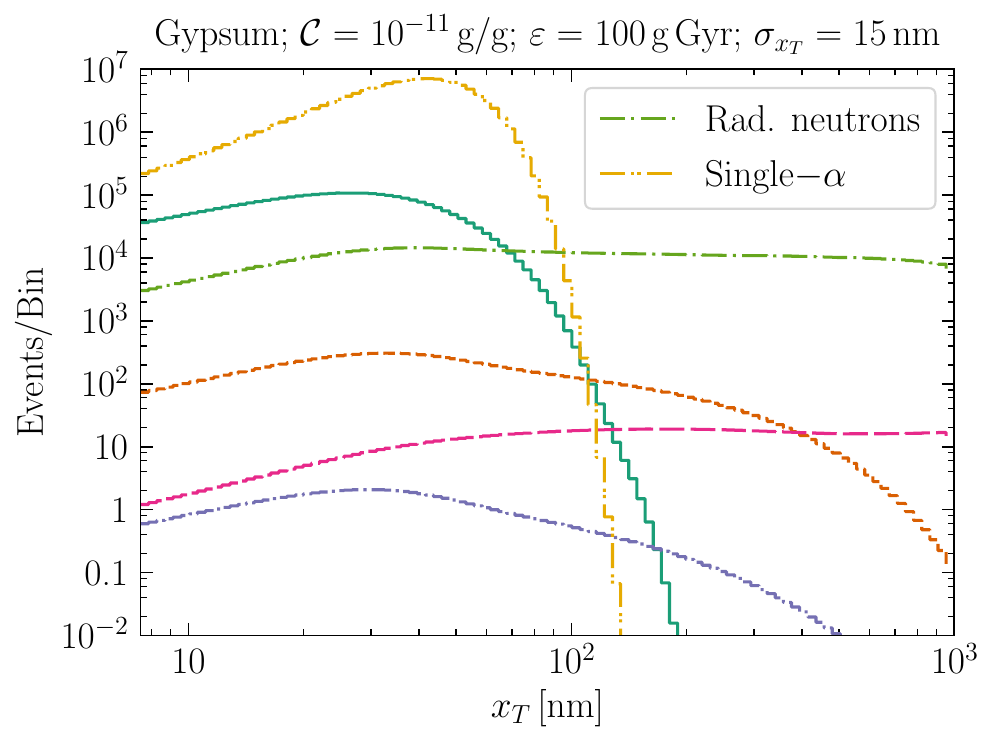}
    \includegraphics[width=0.49\linewidth]{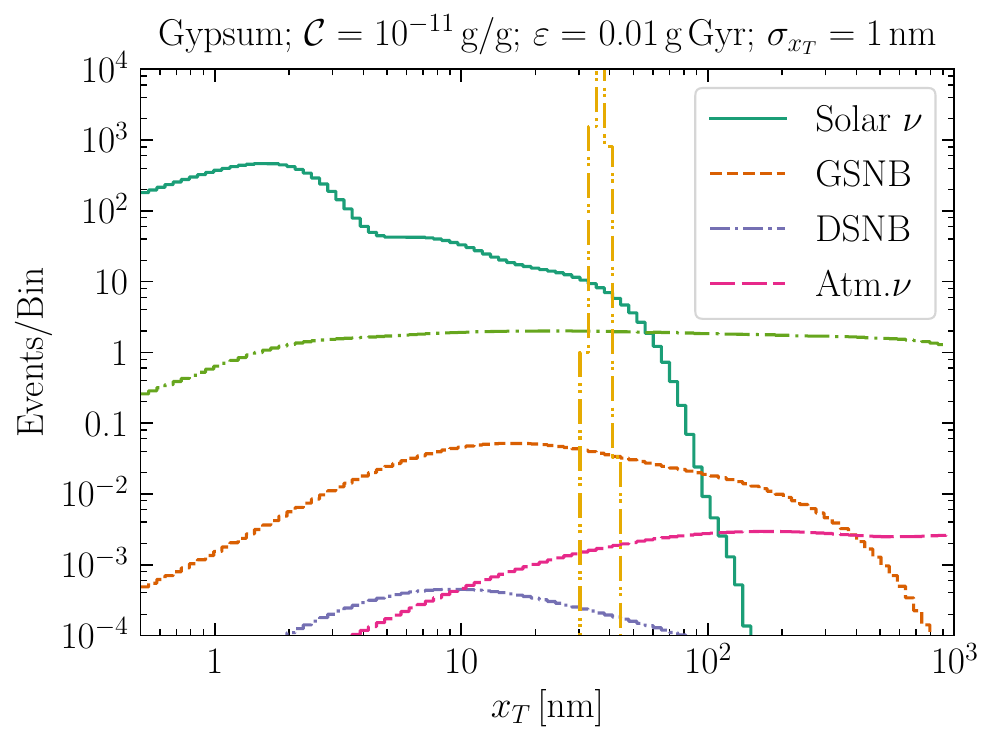}
    \caption{Examples of binned background spectra for the low (left) and high (right) resolution scenarios. Throughout this paper, we use gypsum $\[ \mathrm{Ca(SO_4)}\mathrm{\cdot2(H_2O)} \]$ as the target material, and assume a $^{238}$U concentration of $\mathcal{C} = 10^{-11}\,$g/g. The different lines show the different background contributions discussed in the text: radiogenic (rad.) neutrons, the $^{234}$Th tracks from single-$\alpha$ decays of $^{238}$U, solar neutrinos (solar $\nu$),  Galactic supernova neutrinos (GSNB), diffuse supernova neutrinos (DSNB), and atmospheric neutrinos (atm. $\nu$). In both panels we have used 100 logarithmically spaced bins from $\sigma_{x_T}/2$ to $10^3\,$nm; this binning matches all analyses below. The left panel is for the high-exposure readout scenario, where we assume an exposure of $\varepsilon = 100\,$g\,Gyr and a spatial resolution of $\sigma_{x_T} = 15\,$nm, while the right panel is for the high-resolution readout scenario ($\varepsilon = 0.01\,$g\,Gyr, $\sigma_{x_T} = 1\,$nm). Note that because of the different $\sigma_{x_T}$, the range of the $x$-axis, as well as the width of the bins, differs between the two panels. \textit{Left:} Here, radiogenic backgrounds dominate the background budget for all track lengths. More specifically, for $x_T \lesssim 10^2\,\mathrm{nm}$, the \textit{smeared} single-$\alpha$ tracks are the largest background, whereas for $x_T \gtrsim 10^2\,\mathrm{nm}$, radiogenic neutron induced tracks become dominant. \textit{Right:} For $x_T \gtrsim 10^2\,\mathrm{nm}$, radiogenic neutrons remain the dominant background. Unlike the low resolution scenario, the single-$\alpha$ background is clearly resolved; for $x_T \lesssim 10^2\,\mathrm{nm}$, the dominant background then becomes solar neutrinos.}
    \label{Fig:bkgs}
\end{figure*}

\medskip

\textbf{\textit{Backgrounds ---}} The background sources in paleo-detectors are similar to those in conventional direct detection experiments~\cite{Schumann:2019eaa}: cosmic rays, radioactive decays, and (astrophysical) neutrinos. However, there are quantitative differences in the relative importance of these sources between paleo-detectors and conventional experiments for a number of reasons. First, the exposures of paleo-detectors are much larger than those of conventional direct detection experiments. Thus, unlike conventional direct detection experiments in which one typically tries to construct a signal region with very few (or, ideally, zero) background events, a paleo-detector would contain a large number of background (and, potentially, signal) events. Second, paleo-detectors require only relatively small samples, $M_s < \mathcal{O}(1)\,$kg. Such samples can be obtained from very deep underground, for example, from existing boreholes, providing much better shielding from cosmic ray induced backgrounds than what is attained in existing underground laboratories where conventional detectors are operated. Third, electrons and photons do not produce damage tracks, making paleo-detectors insensitive to electronic recoils.

We will assume that the mineral samples used as paleo-detectors have been shielded from cosmic rays by an overburden of $\gtrsim 5\,$km rock since they started recording nuclear damage tracks --- this is sufficient to suppress cosmogenic background to a negligible level~\cite{Baum:2018tfw, Drukier:2018pdy}.\footnote{Note that the samples can be stored close to the surface for a few years after extraction and prior to readout without accumulating significant cosmogenic backgrounds. For example, the cosmogenic-muon-induced neutron flux in a 50\,m deep storage facility is $\lesssim 0.2\,{\rm cm}^{-2}\,{\rm yr}^{-1}$.} However, there will be a sizable number of neutrino-induced and radiogenic background events in a paleo-detector. In Fig.~\ref{Fig:bkgs}, we show the associated (binned and smeared) track length spectra in the {\it high} (left panel) and {\it low} (right panel) resolution readout scenarios in gypsum $\[ \mathrm{Ca(SO_4)}\mathrm{\cdot2(H_2O)} \]$. 

Neutrinos induce nuclear recoils by scattering off the nuclei in the target mineral. The most relevant neutrino sources for DM searches in paleo-detectors are our Sun, supernovae, and cosmic rays interacting with Earth's atmosphere. We model the neutrino-induced track length spectra as in Refs.~\cite{Baum:2018tfw, Drukier:2018pdy, Baum:2019fqm}, with solar and atmospheric neutrino fluxes taken from Ref.~\cite{OHare:2020lva}. Since the integration times are much longer than the time between supernovae in our Galaxy (approximately 2--3 per century), paleo-detectors would not only record nuclear recoil tracks from the diffuse supernova neutrino background (DSNB), but also those induced by galactic supernovae (the Galactic supernova neutrino background, GSNB). We model the DSNB and GSNB as in Ref.~\cite{Baum:2019fqm}. In this work, we will treat these neutrino fluxes as constant in time, however we account for violations of this and other modeling assumptions via a systematic modeling uncertainty (see Secs.~\ref{sec:uncertainty_stats}/\ref{sec:uncertainty}). Considering the different neutrino-induced background spectra in Fig~\ref{Fig:bkgs}, we see that at short track lengths ($x_T \lesssim 100\,\mathrm{nm}$) solar neutrinos contribute most tracks, at intermediate lengths ($100\,{\rm nm} \lesssim x_T \lesssim 400\,$nm) the GSNB dominates, and for $x_T \gtrsim 400\,$nm atmospheric neutrinos are the largest neutrino-induced background.

Radiogenic backgrounds primarily originate from $^{238}{\rm U}$ and its decay products. While the half-life of $^{238}{\rm U}$ ($T_{1/2} \sim 4\,$Gyr) is long compared to the age of paleo-detector samples, the subsequent decays in the uranium series,
\begin{equation}\begin{split} \label{eq:Uchain}
 &{ ^{238}{\rm U} } \stackrel{\alpha}{\longrightarrow} { ^{234}{\rm Th} } \stackrel{\beta^-}{\longrightarrow} { ^{234{\rm m}}{\rm Pa} } \stackrel{\beta^-}{\longrightarrow} { ^{234}{\rm U}} \stackrel{\alpha}{\longrightarrow} {^{230}{\rm Th}} \\
 & \quad \stackrel{\alpha}{\longrightarrow} {^{226}{\rm Ra}} \stackrel{\alpha}{\longrightarrow} {^{222}{\rm Rn}} \stackrel{\alpha}{\longrightarrow} \ldots \longrightarrow {^{206}{\rm Pb}}\;,
\end{split}\end{equation}
are much faster --- the accumulated half-life of all decays from $^{234}$Th until the stable $^{206}{\rm Pb}$ is $\sim 0.3\,$Myr. Thus, almost all $^{238}{\rm U}$ nuclei which undergo the initial ($^{238}{\rm U} \to {^{234}{\rm Th}} + \alpha$) decay will have completed the uranium series to the stable $^{206}{\rm Pb}$. In an $\alpha$-decay\footnote{$\beta/\gamma$-decays do not give rise to nuclear recoils sufficiently energetic to produce a nuclear damage track.}, the child nucleus recoils with $\mathcal{O}(10-100)\,\mathrm{keV}$ energy and leaves a corresponding track. There are eight $\alpha$-decays in the uranium series; the directions of the associated recoils are uncorrelated and will therefore lead to an interconnected pattern of tracks that is clearly distinguishable from an isolated recoil. We assume that such backgrounds can be completely vetoed during the readout process, although this has yet to be shown in practice.

Unfortunately, the half-life of $^{234}{\rm U}$ (the second $\alpha$-decay in the uranium-series) is relatively long ($T_{1/2} \sim 0.2\,$Myr). Thus, there will be a population of events which have undergone the initial ($^{238}{\rm U} \to {^{234}{\rm Th}} + \alpha$) decay, but not the ($^{234}{\rm U} \to {^{230}{\rm Th}} + \alpha$) decay. These events give rise to isolated tracks from the $72\,\mathrm{keV}$ recoil the $^{234}$Th receives in the $^ {238}$U decay~\cite{Collar:1995aw,SnowdenIfft:1996zz}. For the high-resolution scenario (right panel of Fig.~\ref{Fig:bkgs}), this leads to an almost monochromatic track length spectrum (labeled ``single-$\alpha$'') which has little effect on the sensitivity of paleo-detectors to DM. On the other hand, for the low-resolution scenario (left panel of Fig.~\ref{Fig:bkgs}), the single-$\alpha$ background gets smeared out and becomes the dominant background for track lengths $x_T \lesssim 100\,\mathrm{nm}$.

Additional radiogenic backgrounds stem from fast neutrons produced by spontaneous fission of the nuclei in the uranium series and from $(\alpha, n)$-reactions.\footnote{Depending on the particular chemical composition of any mineral, either spontaneous fission or $(\alpha,n)$-reactions are the dominant source of fast neutrons.} As fast neutrons move through a paleo-detector, they scatter off atomic nuclei, typically losing only a small fraction of their energy in any individual neutron-nucleus interaction. Thus, radiogenic neutrons produce a broad track length spectrum, see the green dot-dashed line in Fig.~\ref{Fig:bkgs}. Importantly, the mean free path of MeV neutrons in typical minerals is a few cm, hence, the multiple tracks produced by the interactions of any particular neutron cannot be correlated with each other. As in previous work, we use \texttt{SOURCES-4A}~\cite{sources4a:1999} to calculate the neutron spectrum from spontaneous fission and $(\alpha,n)$-reactions, taking into account contributions from the entire $^{238}$U decay chain. We then use our own Monte Carlo simulation~\cite{Baum:2018tfw, Drukier:2018pdy} to compute the associated nuclear recoil (and, in turn, track length) spectrum based on neutron-nucleus cross sections tabulated in the \texttt{JANIS4.0} database~\cite{Soppera:2014zsj}.\footnote{This Monte Carlo simulation has recently been validated by comparison with results from \texttt{FLUKA}~\cite{Ferrari:2005zk,Bohlen:2014buj,NUNDIS} for the particular case of halite (NaCl)~\cite{Jordan:2020gxx}.} Fortunately, the neutron background can be suppressed significantly by choosing minerals that contain hydrogen; due to their similar mass, neutrons lose a large fraction of their momentum in a single interaction with hydrogen, moderating the neutrons and suppressing the neutron-induced background. 

Comparing the radiogenic and the neutrino-induced background, for the low-resolution scenario, we see from Fig.~\ref{Fig:bkgs} that radiogenics are the dominant background contribution for the entire range of track lengths considered here. For the high-resolution scenario, the single-$\alpha$ background is well-resolved and therefore has little effect on the sensitivity. The dominant background then becomes solar neutrinos at track lengths $x_T \lesssim 100\,$nm, whereas radiogenic neutrons remain dominant at $x_T \gtrsim 100\,\mathrm{nm}$.

\medskip

\textbf{\textit{Mineral Selection ---}} The selection of target materials for paleo-detectors is largely driven by the backgrounds described above. In particular, the normalization of the radiogenic backgrounds is proportional to the concentration of $^{238}$U in the mineral. Furthermore, in minerals containing hydrogen, neutron-induced backgrounds are strongly suppressed. Two promising classes of radiopure minerals are known as \textit{ultra-basic rocks} and \textit{marine evaporites}; see Ref.~\cite{Baum:2019fqm} for a discussion of the expected concentrations of $^{238}$U in realistic minerals. We will focus on gypsum $\[\mathrm{Ca(SO_4)}\mathrm{\cdot2(H_2O)} \]$, one of the most common marine evaporites. As in previous work on paleo-detectors, we will assume a fiducial $^{238}$U concentration of $\mathcal{C} = 10^{-11}\,$g/g; we will also explore the effect larger or smaller $\mathcal{C}$ would have on the sensitivity. We note that while gypsum is a promising target material, since it is radiopure and contains hydrogen, other minerals may be marginally more sensitive to DM signals --- see Ref.~\cite{Drukier:2018pdy} for a discussion on mineral selection. 

\section{Signal Modeling} \label{sec:signals}
So far, we have described the basic principles of paleo-detectors and the most important background sources. In this section, we discuss the calculation of the recoil spectra for DM signals. To set the stage, we briefly review the calculation of the recoil spectra induced by the DM comprising the (smooth) halo of the Milky Way (MW). We then discuss how to extend this formalism to the spectra induced by a paleo-detector traversing different DM substructures. We remind the reader that we have provided a convenient glossary of symbols and notation in Appendix~\ref{app:notation_table}. 

\subsection{MW halo} \label{sec:MWhalo}

The differential rate per unit target mass of recoils for a DM particle with mass $m_\chi\MW$ elastically scattering off nuclei with mass $m_N$ is given by~\cite{Engel:1991wq,Engel:1992bf,Cerdeno:2010jj}
\begin{equation} \label{eqn:recoil_2}
    \left(\dbd{R}{E_R}\right)\MW = \frac{A^2 F^2}{2} \frac{\SIDDMW}{m_\chi\MW (\mu_{\chi p}\MW)^2} \rho_\chi\MW \eta_\chi\MW(v_{\rm min}) \;,
\end{equation}
where, compared to Eq.~\eqref{eq:dRdx}, we have suppressed the index for the different nuclei comprising the target mineral. In Eq.~\eqref{eqn:recoil_2}, we have assumed standard spin-independent (SI) DM-nucleon interactions with equal couplings to protons and neutrons, parametrized by the zero-momentum-transfer DM-proton cross section, $\SIDDMW$.\footnote{See, for example, Refs.~\cite{Engel:1991wq,Engel:1992bf,Ressell:1993qm,Bednyakov:2004xq,Bednyakov:2006ux} for the analogous expression for isospin-violating SI interactions and spin-dependent DM-nucleon interactions, and Refs.~\cite{Fan:2010gt,Fitzpatrick:2012ix} for more general DM-nucleus interactions.} The factor $A^2$ comes from the coherent enhancement for a nucleus composed of $A$ nucleons. The internal structure of the nucleus is encoded in the form factor $F = F(E_R)$, for which we assume the Helm parametrization~\cite{Helm:1956zz,Lewin:1995rx,Duda:2006uk}, and $\mu_{\chi p}\MW = m_\chi\MW m_p / (m_\chi\MW + m_p)$ is the reduced mass of the DM-proton system with the proton mass $m_p$. The DM distribution in the vicinity of the detector is described by the local DM (mass) density, $\rho_\chi\MW$, and the {\it mean inverse speed}, 
\begin{equation}
    \eta\MW(v_{\rm min}) = \int_{v > v_{\rm min}} \frac{f\MW({\bm v})}{v}\,\mathrm{d}^3{\bm v}\,.
\end{equation}
The integral is over DM velocities ${\bm v}$ in the detector frame, with $v = |{\bm v}|$ and $v_\mathrm{min} = \sqrt{m_N E_R/2(\mu_{\chi N}\MW)^2}$ where $\mu_{\chi N}\MW = m_\chi\MW m_N / (m_\chi\MW + m_N)$. We set the local DM density to $\rho_\chi\MW = 0.3\,$GeV/cm$^3$. For the DM velocity distribution, $f\MW({\bm v})$, we assume a Maxwell-Boltzmann distribution with velocity dispersion $\sigma_v\MW = 166\,$km/s~\cite{Koposov:2009hn}, truncated at the Galactic escape speed $v_{\rm esc}\MW = 550\,$km/s~\cite{Piffl:2013mla} and boosted to the Solar System frame by $v_\odot\MW = 248\,$km/s~\cite{Bovy:2012ba}, as in the Standard Halo Model (SHM)~\cite{Drukier:1986tm,Lewin:1995rx,Freese:2012xd}.\footnote{We do not consider here uncertainties on the speed distribution~\cite{Green:2017odb,Wu:2019nhd,Baxter:2021pqo} or more recently suggested refinements to the SHM~\cite{Evans:2018bqy,Buch:2019aiw}.} Note that the orbital speed of the Earth around the Sun, $v_\oplus \approx 30\,$km/s, is much smaller than $\sigma_v\MW$ and $v_\odot\MW$. Hence, including the motion of the Earth around the Sun in the computation of $\eta\MW$ would only lead to a slight Doppler broadening of the velocity distribution and would not have a considerable effect on the recoil spectra; we neglect this motion for the purposes of the MW signal.

In Fig.~\ref{Fig:sigs}, the solid green lines show the (binned and smeared) track-length spectrum for the MW halo signal in the high- and low-resolution readout scenarios for $m_\chi\MW = 500\,\mathrm{GeV}$ and $\SIDDMW = 5\times10^{-46}\,\mathrm{cm}^{2}$ (a cross section close to current upper limits~\cite{Aprile:2018dbl}). The time dependence of the MW signal is illustrated in Fig.~\ref{Fig:timevar}.

\subsection{Dark Disk} \label{sec:darkdisk}

Let us now discuss how to compute the signal that a component of DM confined in a dark disk would induce in a paleo-detector. The dissipative DM component forming the dark disk would be distinct from the DM particles making up the approximately spherical DM halo of the MW. While we will assume that the DM making up the dark disk does interact with nuclei via standard SI interactions [as in Eq.~\eqref{eqn:recoil_2}], its scattering cross section, $\SIDDdisk$, and mass, $m_\chi\disk$, can be different from those of the MW DM, $\SIDDMW$ and $m_\chi\MW$.

\begin{figure*}[th!]
    \includegraphics[width=\linewidth]{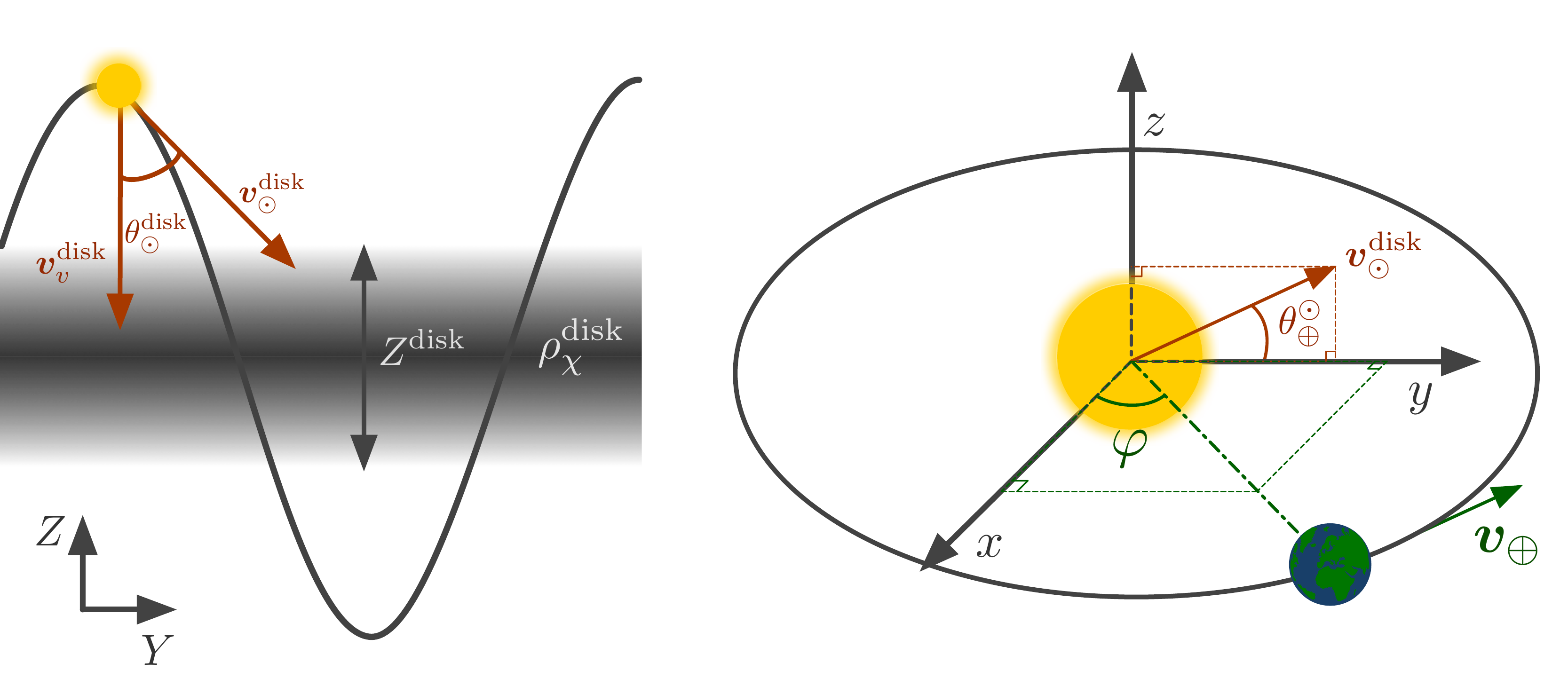}
    \caption{Illustration of the relevant kinematic quantities controlling the relative velocity of a paleo-detector with respect to a dark disk. The Solar System oscillates vertically through the disk with a period of $\sim 90\,\mathrm{Myr}$, as shown in the left panel. In the right panel, we have chosen the Earth's orbit to lie in the $x-y$ plane and the velocity of the Solar System with respect to the disk to lie in the $y-z$ plane. Note that the coordinate axes in the two panels are unrelated. As discussed in the main text, the precise details of each disk crossing differ slightly due to the relative orientation of the Earth with respect to the disk; these details can be found in Table~\ref{tab:crossings}.}
    \label{fig:DD_coordinates}
\end{figure*}

There are three additional important differences between the track length spectra induced by a dark disk and by the MW halo. First, the Solar System passes through the Galactic plane with a vertical velocity of $v\disk_v \sim 7\,{\rm km/s} \sim 7\,{\rm pc/Myr}$~\cite{Schoenrich:2009bx}, and would therefore traverse a dark disk with a thickness $z^{\rm disk} \lesssim 10\,$pc in $\lesssim 1.5\,$Myr (see left panel of Fig.~\ref{fig:DD_coordinates} for an illustration of the Solar System's motion with respect to a dark disk). Thus, the duration of a disk-crossing is short compared to the age of paleo-detector samples, $T^n \sim 10\,{\rm Myr} - 1\,$Gyr, see Fig.~\ref{Fig:timevar}. Second, since the relative speed of the Solar System with respect to the dark disk is small ($v_\odot\disk \sim 30\,$km/s), and the internal velocity dispersion of the DM making up the dark disk is even smaller~\cite{Fan:2013yva} ($\sigma_v\disk \ll 10\,$km/s), the nuclear recoils induced by DM in a dark disk will be less energetic, and in turn, the recoil tracks much shorter than those induced by DM in the MW halo. Third, because $\sigma_v\disk$ and $v_\odot\disk$ are comparable to the orbital speed of the Earth around the Sun ($v_\oplus \approx 30\,$km/s), we cannot neglect the orbital motion of the Earth when computing the signal from a dark disk. 

Because the time it takes the Solar System to cross the dark disk is small compared to the exposure time of a paleo-detector, it is useful to compute the differential number of recoils induced by crossing through the dark disk once by integrating from the time when the Solar System enters the dark disk ($t_0$) to the time when it leaves ($t_1$), yielding $\dd n/\dd E_R = {\textstyle \int}_{t_0}^{t_1} \dd R/\dd E_R\,\dd t$. If we assume that the velocity of the Solar System with respect to the dark disk, ${\bm v}_\odot\disk$, is constant during $t_0 \leq t \leq t_1$, we can compute $\dd n / \dd E_R$ [see Eq.~\eqref{eqn:recoil_2}],
\begin{widetext}
    \begin{align} \label{eq:disk_single0}
        \left(\dbd{n}{E_R}\right)\disk &= \frac{A^2 F^2}{2} \frac{\SIDDdisk}{m_\chi\disk (\mu_{\chi p}\disk)^2} \int_{t_0}^{t_1} \rho_\chi\disk(t) \eta_\chi\disk(t; v_{\rm min}) \, \dd t \\
        &= \frac{A^2 F^2}{2} \frac{\SIDDdisk}{m_\chi\disk (\mu_{\chi p}\disk)^2} \frac{\cos\theta_\odot\disk}{v_v\disk} \int_{-Z\disk/\cos\theta_\odot\disk}^{Z\disk/\cos\theta_\odot\disk} \rho_\chi\disk(\ell) \eta_\chi\disk(\ell; v_{\rm min}) \, \dd\ell \\
        &= \frac{A^2 F^2}{2} \frac{\SIDDdisk}{m_\chi\disk (\mu_{\chi p}\disk)^2} \frac{\Sigma\disk}{v_v\disk} \overline{\eta}_\chi\disk(v_{\rm min})\;, \label{eq:disk_single}
    \end{align}
\end{widetext}
where $\theta_\odot\disk$ denotes the angle between $\bm{v}_\odot\disk$ and $\bm{v}_v\disk$, $\Sigma\disk = {\textstyle \int} \rho_\chi\disk(Z)\,\dd Z$ is the surface density of the dark disk, and $\overline{\eta}_\chi\disk$ denotes the mean inverse speed averaged over the crossing.

As mentioned previously, since $v_\oplus\gtrsim\sigma_v\disk,v_\odot\disk$, we must account for the orbit of the Earth around the Sun when evaluating $\overline{\eta}_\chi\disk$. We will assume that this orbit is circular and work in Cartesian coordinates with the Earth's orbit lying in the $x-y$ plane. We denote the phase of Earth's orbit around the Sun with $\varphi$ and the orbital velocity of Earth with ${\bm v}_\oplus$. We furthermore choose the velocity of the Sun with respect to the disk, ${\bm v}_\odot\disk$, to lie in the $y-z$ plane,\footnote{Note that the $z$ coordinate denoting the direction perpendicular to the Earth's orbit is distinct from the $Z$ coordinate perpendicular to the galactic disk.} and denote the angle between ${\bm v}_\odot\disk$ and the orbital plane with $\theta_\oplus^\odot$.\footnote{We will assume that ${\bm v}_\oplus$ and ${\bm v}_\odot\disk$ are constant for the duration of the disk crossing.} These coordinates are best understood visually, see Fig.~\ref{fig:DD_coordinates}. The relative velocity of Earth with respect to the dark disk is then 
\begin{equation}
    {\bm v}_{\rm rel}\disk = v_\oplus \begin{pmatrix} -\sin\varphi \\ \cos\varphi \\ 0 \end{pmatrix} + v_\odot\disk \begin{pmatrix} 0 \\ \cos\theta_\oplus^\odot \\ \sin\theta_\oplus^\odot \end{pmatrix} \;,
\end{equation}
and its magnitude is
\begin{align}
    v_{\rm rel}\disk(\varphi) = \sqrt{2 v_\oplus v_v\disk \frac{\cos\theta_\oplus^\odot}{\cos\theta_\odot\disk} \cos\varphi + v_\oplus^2 + \frac{(v_v\disk)^2}{\cos^2\theta_\odot\disk}} \;.
\end{align}
The time-averaged mean inverse speed is thus given by
\begin{equation}
    \overline{\eta}\,\disk(v_{\rm min}) = \frac{1}{2\pi} \int_0^{2\pi} \dd \varphi\; \int_{v > v_{\rm min}} \dd^3 {\bm v}\; \frac{\tilde{f}\disk({\bm v}-{\bm v}_{\rm rel}\disk)}{v_{\rm rel}\disk} \;,
\end{equation}
where $\tilde{f}\disk$ is the velocity distribution of the DM in the rest frame of the dark disk. For concreteness, we will assume a Maxwell-Boltzmann distribution for $\tilde{f}\disk$ with velocity dispersion $\sigma_v\disk = 1\,$km/s.\footnote{We will assume $\sigma_v\disk = 1\,$km/s throughout. As long as $\sigma_v^{\rm disk}$ is small compared to $v_v\disk$ and $v_\oplus$, the effect of changing $\sigma_v\disk$ on the induced nuclear recoil spectrum is negligible, especially after taking into account finite resolution effects. Furthermore, since $\sigma_v\disk \ll v_{\rm rel}\disk$, the precise form of $\tilde{f}(v)\disk$ has virtually no effect on the dark disk induced signal.}

\begin{table}
    \renewcommand\arraystretch{1.4}
    \centering
    \begin{tabular}{C{2cm} C{2cm} C{2cm} C{2cm}}
        \hline\hline
        $t_i\disk$ [Myr] & $v_v\disk$ [km/s] & $\theta_\odot\disk~[^{\circ}]$ & $\theta_\oplus^\odot~[^{\circ}]$ \\ 
        \hline
        2.3 & 7.4 & 103 & 31\\
        43.5 & $-7.2$ & $-109$ & 4.4\\
        89.8 & 6.8 & 109 & $-22$\\
        139 & $-6.9$ & $-111$ & $-27$\\
        183 & 7.4 & 102 & 2.1\\
        224 & $-7.2$ & $-109$ & 29\\
        \hline\hline
    \end{tabular}
    \caption{Parameters of crossings of the Galactic mid-plane in the past 250 million years, where $t_i\disk$ is the time \textit{before present} at which the crossing occurred, and $v_v\disk$, $\theta_\odot\disk$, and $\theta_\oplus^\odot$ are the kinematic parameters defined in Fig.~\ref{fig:DD_coordinates}.}
    \label{tab:crossings}
\end{table}

In Eq.~\eqref{eq:disk_single}, we have computed the signal from a single disk crossing. The signal in a paleo-detector sample which has been recording tracks for a time $T$ is then given by summing Eq.~\eqref{eq:disk_single} over all disk crossings which occurred at times $t_i\disk \leq T$ before the present. Importantly, each disk crossing has slightly different kinematics. We use \texttt{galpy}~\cite{2015ApJS..216...29B} to simulate the Solar System's orbit through the Galaxy in order to compute these kinematic parameters\footnote{We adopt the \texttt{MWPotential2014} Galactic potential, which has been fit to a variety of existing measurements (see Section~3.5 of Ref.~\cite{2015ApJS..216...29B} for a discussion). We set the Galactocentric radius of the Solar System to $R_{\odot} = 8\,$kpc, the present height of the Sun above the Galactic plane to $z = 17.4\,$pc~\cite{2017MNRAS.465..472K}, and the present velocity of the Sun in Galactocentric coordinates to $(v_R, v_T, v_z) = (12.24, 231.1, 7.25)\,$km/s~\cite{Schoenrich:2009bx}. The angle between the ecliptic plane and the Galactic plane is fixed at $30^{\circ}$ with the $R-\phi$ projection of the ecliptic pole oriented in the tangential direction. With these parameters, we simulate the orbit in reverse to compute the times at which the Sun crossed the Galactic plane and the associated kinematic quantities at each crossing. We have also manually adjusted these parameters to assess the dependence of our results on this particular choice and find that they are very insensitive to $\mathcal{O}(1)$ changes in these parameters.} and list them in Table~\ref{tab:crossings}.

\begin{figure*}[ht!]
    \includegraphics[width=0.49\linewidth]{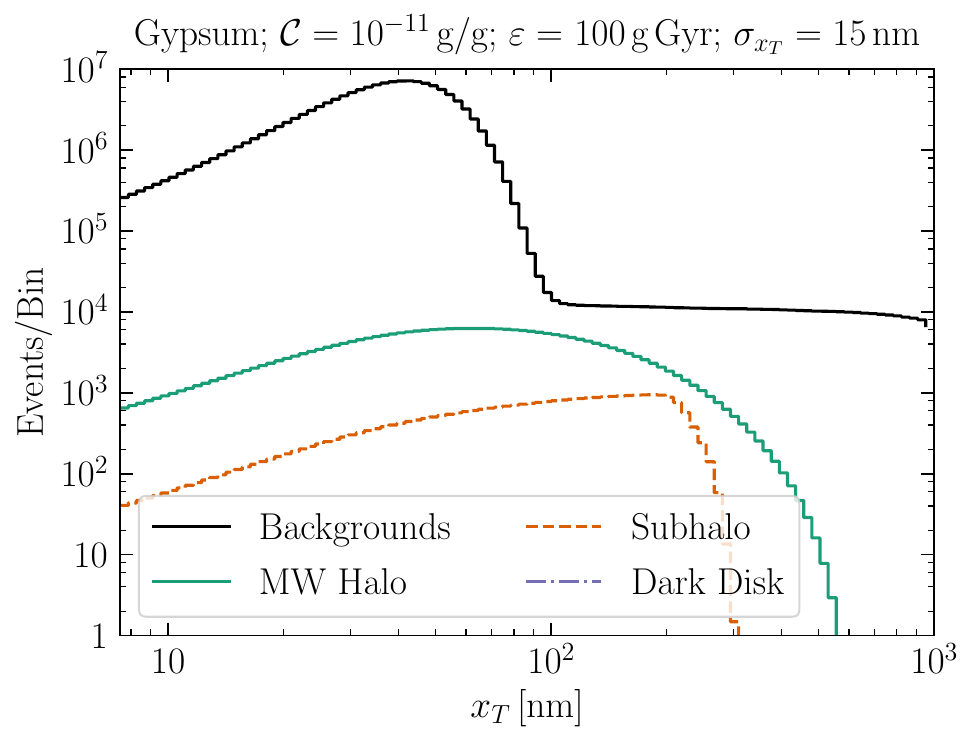}
    \includegraphics[width=0.49\linewidth]{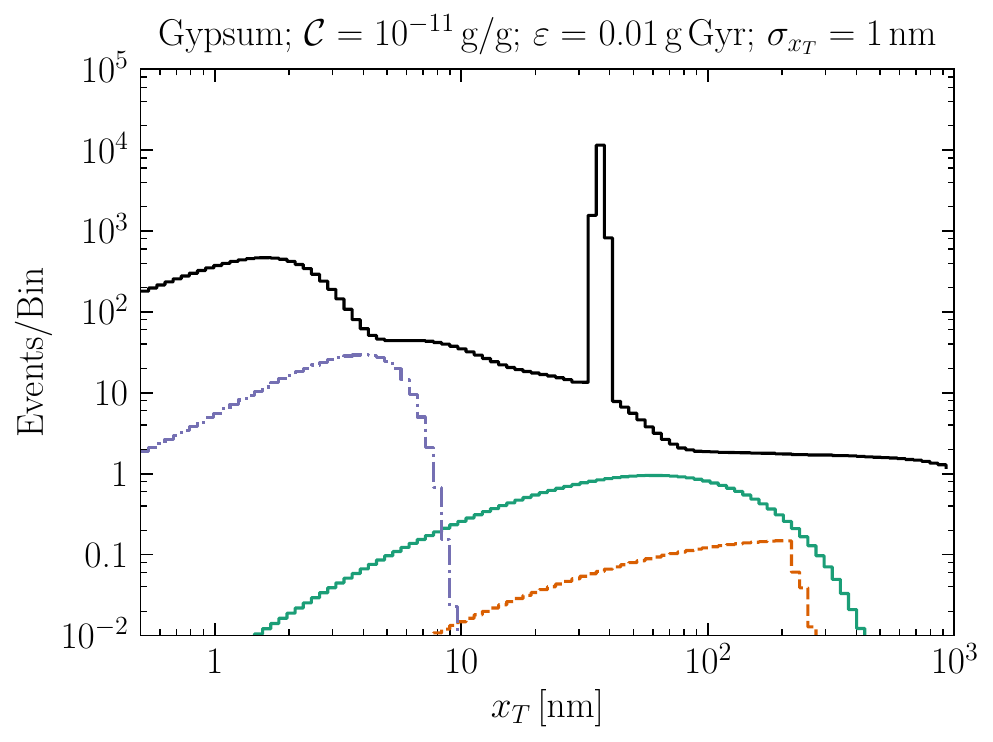}
    \caption{Examples of binned signal spectra, with summed backgrounds from Fig.~\ref{Fig:bkgs}. We use 100 logarithmically spaced bins from $\sigma_{x_T}/2$ to $10^3\,$nm. Note that due to the different $\sigma_{x_T}$ in the high-exposure (left) and high-resolution (right) readout scenarios, the range of the $x$-axes differs between the two panels. For the dark disk signal, we include a single crossing within the age of the sample, using the kinematic parameters from the first line of Table~\ref{tab:crossings}, and fix the DM parameters to be $m_\chi\disk = 100\,\mathrm{GeV}$, $\SIDDdisk = 10^{-43}\,\mathrm{cm}^{2}$, and $\Sigma\disk = 10\,M_\odot/\mathrm{pc}^2$; a $\mathcal{ O}(1)\,$Gyr-old sample would measure the sum of the spectra from $\mathcal{O}(20)$ such crossings, each with different kinetic parameters. For the subhalo signal, we use $m_\chi\SH = 500\,\mathrm{GeV}$ and $\SIDDSH = 5\times10^{-46}\,\mathrm{cm}^{2}$ for the DM particle parameters and fix the subhalo parameters to: $M_{\rm vir}\SH=10^6\,M_\odot$; $c\SH=65$; $v_\odot\SH = 500 \,\mathrm{km/s}$; and $b\SH/r\SH_s= 10^{-2}$. The MW halo background uses the same DM particle parameters as the subhalo signal.}
    \label{Fig:sigs}
\end{figure*}

In Fig.~\ref{Fig:sigs} we show the track length spectrum induced by a {\it single}\footnote{We use the kinematic parameters from the first line of Table~\ref{tab:crossings} for definiteness.} dark disk crossing in a paleo-detector (purple dash-dotted line) for $m_\chi\disk = 100\,\mathrm{GeV}$, $\SIDDdisk = 10^{-43}\,\mathrm{cm}^{2}$, and $\Sigma\disk = 10\,M_\odot/\mathrm{pc}^2$ (see also Fig.~\ref{Fig:timevar} for an illustration of the time dependence of the dark disk signal). Note that, as discussed previously, the tracks produced by DM in a dark disk are much shorter than those from the MW halo (shown by the green solid line in Fig.~\ref{Fig:sigs}) because the relative speed of the Solar System with respect to the rest frame of the dark disk, $v_\odot\disk$, is much smaller than the relative speed of the Solar System with respect to the MW halo rest frame, $v_\odot\MW$. In particular, the tracks induced by crossing the dark disk are so short for this particular choice of parameters that the corresponding track length spectrum does not appear in the left panel of Fig.~\ref{Fig:sigs}, which shows the low-resolution scenario.\footnote{This is due to our conservative cut, removing all tracks with true length $x_T' < \sigma_{x_T}/2$.} Thus, we can already see that the high-resolution readout scenario is much better suited to searching for a signal induced by a dark disk than the low-resolution scenario.

\subsection{Subhalo} \label{sec:subhalo}

Let us consider subhalos formed from the same DM particles as those comprising the MW halo; in this case, the particle masses and scattering cross sections are the same ($m_\chi\SH = m_\chi\MW$ and $\SIDDSH = \SIDDMW$). The signal in a paleo-detector induced by traversing a subhalo differs from that induced by the MW halo in two primary ways. First, the DM within the subhalo has much smaller velocity dispersion than the DM in the MW halo~\cite{BinneyTremaine:1987}, $\sigma_v\SH \ll \sigma_v\MW$, such that the subhalo appears as an approximately monochromatic \textit{wind} of DM particles with their velocity set by the relative motion of the Solar System with respect to the subhalo, $v_\odot\SH$.\footnote{Note that since $\sigma_v\SH \ll v_\odot\SH$, the precise form of the DM speed distribution in the subhalo has virtually no effect on the subhalo signal.} Second, the signal is transient since the Solar System traverses a subhalo on short timescales compared to a paleo-detector's exposure time, see Fig.~\ref{Fig:timevar}. It is the combination of these two features that allow paleo-detectors to be sensitive to a collision with a subhalo. Astronomical observations constrain the halo mass function down to virial masses $M_{\rm vir}\SH \sim \mathcal{O}(10^7)\,M_\odot$~\cite{Nadler:2019zrb,Schutz:2020jox,Nadler:2020prv,Mao:2020rga,Das:2020nwc,Maamari:2020aqz,Nadler:2021dft}. Trivially, heavier subhalos would give rise to larger integrated signals (for the same concentration and impact parameter). Hence, we are mainly interested in the subhalo mass range $M_{\rm vir}\SH \sim 10^4 - 10^8\,M_\odot$ that lies just below current constraints. Regarding the speed of subhalos relative to the Solar System, $v_\odot\SH$, subhalos are expected to follow the same velocity distribution as the DM making up the MW halo, see Sec.~\ref{sec:MWhalo}. Thus, the speed distribution $f(v_\odot\SH)$ would peak at $v_\odot\SH$ of a few hundred km/s, with the maximal encounter speed set by the local Galactic escape speed, $v_\odot\SH \leq v_{\rm esc}\MW + v_\odot\MW \approx 800\,$km/s. Finally, we will be interested in subhalo encounters with impact parameters $b\SH$ less than the scale radius of the subhalo, so we are able to probe the dense inner part of the subhalo.  We remind the reader that while subhalo encounters with such impact parameters are expected to be rare under standard assumptions of the mass function (see Appendix~\ref{app:subhalo}), they can become more likely with enhanced mass functions from nonstandard cosmologies.

We model the DM density of the subhalo as a Navarro-Frenk-White (NFW) profile~\cite{Navarro:1995iw},
\begin{equation} \label{eq:NFW_dens}
   \rho_\chi\SH(r) = \frac{\rho_s\SH}{r/r_s\SH \left(1 + r/r_s\SH \right)^2} \;,
\end{equation}
where $r$ is the distance from the center of the subhalo, $\rho_s\SH$ the characteristic density, and $r_s\SH$ the scale radius. We parametrize the NFW profile in terms of $M_{\rm vir}\SH$ and the concentration parameter, $c\SH$, of the subhalo. Note that tidal stripping in the MW leads to a truncated NFW profile~\cite{Hiroshima:2018kfv,Ando:2019xlm}. However, the signal from traversing a subhalo is dominated by the dense central region, hence, neglecting this truncation will not affect our results. To compute $\rho_s\SH$ and $r_s\SH$ as functions of $M_{\rm vir}\SH$ and $c\SH$ we follow Ref.~\cite{Ando:2019xlm}. The virial radius is given by
\begin{equation}\label{eq:NFW_virradius}
   r_{\rm vir}\SH(M_{\rm vir}\SH, z) = \left( \frac{3}{4\pi} \frac{M_{\rm vir}\SH}{\Delta_{\rm vir}(z) \rho_c(z)} \right)^{1/3} \;,
\end{equation} 
where $\Delta_{\rm vir}(z)$ is the critical overdensity required for a subhalo to decouple from the cosmic expansion (see Ref.~\cite{Ando:2019xlm}) and $\rho_c(z)$ is the critical density at redshift $z$. The scale radius is related to the virial radius via $ r_s\SH = r\SH_{\rm vir}/c\SH$ and the characteristic density is given by
\begin{equation}\label{eq:NFW_rhos}
   \rho_s\SH(M_{\rm vir}\SH, c\SH, z) = \frac{(c\SH)^3}{f(c\SH)} \Delta_{\rm vir}(z) \rho_c(z) \;,
\end{equation}
where $f(c) = \ln(1+c) - c/(1+c)$. Note that any change in $z$ can be compensated for by a change in $c\SH$. We will therefore fix $z = 0$ and parametrize the density of a subhalo solely by the concentration parameter. 

Even for a large subhalo with $M_{\rm vir}\SH = 10^8\,M_\odot$, the vast majority of the signal would be accumulated within the central few hundred pc. Hence, for a relative speed of the Solar System relative to the subhalo of $v_\odot\SH \sim 100\,{\rm km/s} \sim 100\,{\rm pc/Myr}$, the signal would be accumulated within a few Myr, much shorter than the integration time of a paleo-detector. Thus, we are interested in the differential number of recoils, $\dd n/\dd E_R = {\textstyle \int}_{t_0}^{t_1} \dd R/\dd E_R\,\dd t$, induced by crossing a subhalo where the integral is over the time spent within the subhalo. We will neglect the gravitational attraction from the subhalo and treat the Solar System as traversing the subhalo along a straight line. The encounter is then parametrized by the impact parameter, $b\SH$, of the Solar System relative to the center of the subhalo, the relative speed, $v_\odot\SH$, and how long ago the Solar System was closest to the center of the subhalo, $T\SH$. Performing a calculation analogous to Eqs.~\eqref{eq:disk_single0}--\eqref{eq:disk_single} yields
\begin{equation} \label{eq:dndE_SH} \begin{split}
    \left(\dbd{n}{E_R}\right)\SH &= \frac{A^2 F^2}{2} \frac{\SIDDSH}{m_\chi\SH (\mu_{\chi p}\SH)^2} \\
    &\quad \times \frac{1}{v_\odot\SH} \int_{x_0}^{x_1} \rho_\chi\SH(r) \eta_\chi\SH(r; v_{\rm min}) \, \dd x \;,
\end{split} \end{equation}
where $x$ denotes the trajectory of the Solar System through the subhalo and $r(x) = \sqrt{(b\SH)^2 + x^2}$.

For the velocity distribution of the DM making up the subhalo, we will assume a Maxwell-Boltzmann distribution boosted to the Solar System frame by $v_\odot\SH$. Assuming a virialized subhalo, the velocity dispersion and escape speed can be analytically calculated as a function of $r$~\cite{BinneyTremaine:1987}. The motion of the Earth around the Sun leads to a broadening of the associated recoil spectrum, however, this effect will be negligible unless $\max\left[ v_\odot\SH, \sigma_v\SH(r) \right] \lesssim v_\oplus$; we will neglect this motion in our computation of the recoil spectra.

In Fig.~\ref{Fig:sigs}, the dashed orange lines show the track length spectrum induced by crossing a $M_{\rm vir}\SH = 10^6\,M_\odot$, $c\SH = 65$ subhalo with impact parameter $b\SH=10^{-2}\,r_s\SH$ and velocity $v_\odot\SH = 500 \,\mathrm{km\,s}^{-1}$, assuming the DM is comprised of particles with mass $m_\chi\SH = 500\,\mathrm{GeV}$ and scattering cross section $\SIDDSH = 5\times10^{-46}\,\mathrm{cm}^{2}$ (see also Fig.~\ref{Fig:timevar} for an illustration of the time dependence of the subhalo signal). For these parameters, crossing the subhalo gives rise to fewer tracks than the MW halo signal, however, this hierarchy can be reversed for larger $M_{\rm vir}\SH$ or $c\SH$, or smaller $b\SH$.\footnote{Note that this choice of $M_{\rm vir}\SH$ and $c\SH$ was informed by the mass-concentration relation from Ref.~\cite{Moline:2016pbm}, and this relation predicts that $c\SH$ decreases as $M_{\rm vir}\SH$ increases.  Moreover the choice of $b\SH$ here is already very small compared to $r_s\SH$.  For these reasons, we expect the hierarchy shown in Fig.~\ref{Fig:sigs} to be the more frequent one.}

Note that this subhalo analysis can be straightforwardly extended to a signal from a DM stream, as discussed in Refs.~\cite{Evans:2018bqy, OHare:2018trr}. Similarly to subhalos, a DM stream would have a large relative velocity along the direction of the encounter, but small velocity dispersion. We leave a dedicated analysis of the sensitivity of paleo-detectors to DM streams to future work.

\section{Sensitivity} \label{sec:stat}

In this section, we describe the statistical framework we use to calculate the projected sensitivity of a series of paleo-detectors of different ages to DM substructure. We use a standard profile likelihood ratio approach to perform nested model comparison. Previous work~\cite{Baum:2018tfw, Drukier:2018pdy, Edwards:2018hcf,Baum:2021jak} was mostly interested in the sensitivity of paleo-detectors to the MW halo DM signal amidst various backgrounds. In this paper, we are instead interested in trying to distinguish a time-varying signal from the time-invariant signal that the MW halo would induce. 

The general strategy is as follows: for both the dark disk and the subhalo scenarios, we parametrize the signal with a parameter that approximately controls its overall normalization. For the dark disk, we use the product of the scattering cross section and the surface density, $\SIDDdisk \Sigma\disk$, while for the subhalo case, we use the impact parameter, $b\SH$. Larger values of $\SIDDdisk \Sigma\disk$ correspond to a larger disk signal, while smaller values of $b\SH$ correspond to a larger subhalo signal. To estimate the sensitivity, we compute the smallest (largest) value of $\SIDDdisk \Sigma\disk$ ($b\SH$) for which the time-varying signal+MW halo+backgrounds hypothesis would be preferred over the MW halo+backgrounds-only hypothesis, holding all other parameters controlling the signal fixed. We define the \emph{discrimination reach} as the value of $\SIDDdisk \Sigma\disk$ or $b\SH$ for which 50\,\% of experiments would find a preference for substructure at 95\,\% confidence level (or approximately $2\,\sigma$)~\cite{2012PhRvD..85c5006B, Cowan:2010js}.

For the dark disk scenario, the DM component that makes up the dark disk and the component comprising the MW halo are distinct. Therefore, when computing the discrimination reach, we will assume that the particles making up the smooth MW halo do not give rise to a measurable signal in paleo-detectors, i.e., we use mock data sets generated for a true value of $\SIDDMW = 0$. On the other hand, for the subhalo case, we will assume that the same DM particles make up both the MW-halo and the subhalo. Accordingly, we use mock data sets containing signals from both the subhalo and the MW halo, setting $\SIDDSH = \SIDDMW$ and $m_\chi\SH = m_\chi\MW$.

In order to explain the statistical treatment in more detail, let us start by defining the likelihood function. (Note that we have provided a convenient glossary of the symbols introduced in the following discussion in Appendix~\ref{app:notation_table}.) We consider a series of paleo-detectors (indexed by $n$) with different ages, $T^n$. We use the index $i$ for the different track-length bins (in the $n$-th sample).\footnote{Throughout this paper we use 100 logarithmically spaced bins from $\sigma_{x_T}/2$ to $10^3\,$nm.} We denote the parameters controlling the dark disk/subhalo signal by $(\zeta_0, {\bm \zeta})$, where $\zeta_0$ is the parameter we use to parametrize the normalization of the signal ($\zeta_0 = \SIDDdisk \Sigma\disk$ for the dark disk scenario and $\zeta_0 = b\SH$ for the subhalo scenario) while ${\bm \zeta}$ is the remaining set of parameters controlling the dark disk/subhalo signal. The log-likelihood to observe the data set ${\bm \DD}$ (with entry $\DD_i^n$ in the $i$-th bin of the $n$-th sample) for a set of nuisance parameters ${\bm \theta}$ and parameters $(\zeta_0, {\bm \zeta})$ is\footnote{Here and in the following, we drop constant factors in the expression of the likelihood which cancel in the likelihood ratio we are ultimately interested in.} 
\begin{equation} \label{eqn:likelihood} \begin{split}
    \ln\LL\left( {\bm \DD} \middle| {\bm \theta}; \zeta_0, {\bm \zeta} \right) &= \sum_n \sum_i \left[ \DD_i^n \ln N_i^n - N_i^n \right] \\
    &\quad - \frac{1}{2} \sum_j \left( \frac{\theta_j - \overline{\theta}_j}{c_j \overline{\theta}_j} \right)^2 \;,
\end{split} \end{equation}
 where $N_i^n({\bm \theta}; \zeta_0, {\bm \zeta})$ denotes the expected number of tracks (in the $i$-th bin of the $n$-th sample) for a given set of parameters $({\bm \theta}; \zeta_0, {\bm \zeta})$. Since paleo-detectors are ultimately counting experiments, we expect the data to be Poisson-distributed, corresponding to the contribution in the first line of Eq.~\eqref{eqn:likelihood}. The second line accounts for external constraints on a subset of the nuisance parameters. In frequentist terms, these Gaussian constraints mimic the effect of performing a joint analysis in order to incorporate ancillary measurements of the nuisance parameters. In Eq.~\eqref{eqn:likelihood}, $\overline{\theta}_j$ is the central value of the $j$-th nuisance parameter inferred from an ancillary measurement, and $c_j$ is the associated relative ($c_j\bar\theta_j$ the absolute) uncertainty. 

The set of nuisance parameters we consider is
\begin{equation}
    {\bm \theta} = \left\{ T^n, M_s^n, \mathcal{C}^n, \Phi^{\bm \nu}, \SIDDMW, m_\chi\MW \right\} \;.
\end{equation}
Here, $T^n$, $M_s^n$, and $\mathcal{C}^n$ are the age\footnote{Recall that we use ``age'' to refer to the time a mineral has been recording nuclear damage tracks.}, mass, and $^{238}$U concentration of the $n$-th sample. The $\Phi^{\bm \nu}$ are the fluxes of the various neutrino backgrounds, ${\bm \nu} = \{ {\rm solar}~\nu, \, {\rm GSNB}, \, {\rm DSNB} , \, {\rm atm.}~\nu \}$, see Sec.~\ref{sec:basics} and Fig.~\ref{Fig:bkgs}. 

In our fiducial analysis, we will include constraints on the $T^n$, $M_s^n$, $\mathcal{C}^n$ and the $\Phi^{\bm \nu}$ with uncertainties $c_{T^n} = 5\,\%$, $c_{M_s^n} = 0.1\,\%$, $c_{\mathcal{C}^n} = 10\,\%$ and $c_{\Phi^{\bm \nu}} = 100\,\%$. These choices represent our assumptions on how well these parameters could be constrained by ancillary measurements.\footnote{The choice $c_{\Phi^{\bm \nu}} = 100\,\%$ is motivated by the fact that neutrino-induced backgrounds could potentially vary by an $\mathcal{O}(1)$ factor over $\sim\,$Gyr~\cite{Baum:2019fqm,Jordan:2020gxx,Arellano:2021jul}. Radiogenic backgrounds, on the other hand, do not vary with time and are only controlled by $\mathcal{C}$. Through a combination of direct $\mathcal{C}$ measurements in samples~\cite{Povinec:2018,Povinec:2018wgd} and calibration studies with high-$\mathcal{C}$ samples, the shape and normalization of the radiogenic-induced background can be measured; we therefore assign $c_{\mathcal{C}^n} = 10\,\%$. Mineral samples can be dated to few-percent accuracy using geological dating techniques~\cite{GTS2012,Gallagher:1998,vandenHaute:1998}, therefore motivating $c_{T^n} = 5\,\%$. Finally, although the mass of the sample can be measured precisely, the total sensitive volume will have some uncertainty due to tracks close to the boundaries; we therefore assign $c_{M_s^n} = 0.1\,\%$. We do not include constraints on $\SIDDMW$ and $m_\chi\MW$ in our analysis.} However, as we will see in Sec.~\ref{sec:results}, our results have very little dependence on these choices.

In Eq.~\eqref{eqn:likelihood}, ${\bm N}({\bm \theta}; \zeta_0, {\bm \zeta})$ (with entries $N_i^n$) denotes the expected number of tracks after binning and smearing, see Eq.~\eqref{eq:Ni}. In particular, ${\bm N}$ is the sum of the spectra for the various backgrounds and the relevant signal. The contributions from the backgrounds and from the DM making up the MW halo, ${\bm N}_0({\bm \theta})$, are
\begin{equation} \label{eqn:nullspec} \begin{split}
    (N_0)_i^n({\bm \theta}) &= M^n \left[ T^n \sum_j R_i^{\nu_j} (\Phi^{\nu_j}) \right. \\
    & \qquad \left. \quad~ + n^{1\alpha}_i(\mathcal{C}^n) + T^n R^{\rm neu}_i(\mathcal{C}^n) \right. \\
    & \qquad \left. \phantom{\sum_{{\bm \nu}_j}} + T^n R\MW_{i}(\SIDDMW, m_\chi\MW) \right] \;.
\end{split} \end{equation}
The first line is the contribution from the respective neutrino backgrounds, the second line denotes the radiogenic backgrounds, separated into the ``single-$\alpha$'' (1$\alpha$) and the radiogenic neutron (neu) backgrounds, and the third line is the contribution induced by DM in the MW halo. For all contributions except for the single-$\alpha$ background, the number of tracks produced in a sample is proportional to the age of the sample, $T^n$, and accordingly, they enter Eq.~\eqref{eqn:nullspec} via the rate (per unit target mass) at which tracks are produced in a given bin, $R_i$, see Eq.~\eqref{eqn:smearing}. For the single-$\alpha$ background, on the other hand, the number of tracks is independent of the age of the sample,\footnote{This holds, to good approximation, for $T_{1/2}^{^{234}{\rm U}} < T^n < T_{1/2}^{^{238}{\rm U}}$, where $T_{1/2}^{^{234}{\rm U}} = 0.25\,$Myr and $T_{1/2}^{^{238}{\rm U}} = 4.5\,$Gyr are the half-lives of $^{234}$U and $^{238}$U, respectively.} such that this contribution enters Eq.~\eqref{eqn:nullspec} via $n_i$, the number of tracks per unit target mass in the $i$-th bin, see Eq.~\eqref{eq:Ni}. 

For the dark disk scenario, the expected number of tracks entering Eq.~\eqref{eqn:likelihood} is then
\begin{equation} \label{eq:Ndisk}
    N_i^n({\bm \theta}; \zeta_0, \bm\zeta) = (N_0)_i^n({\bm \theta}) + M^n n_i\disk(\zeta_0, {\bm \zeta}; T^n) \;, 
\end{equation}
where $\zeta_0 = \SIDDdisk \Sigma\disk$, ${\bm \zeta} = \{ m_\chi\disk \}$ and $n_i\disk$ is the (smeared and binned) signal from a dark disk in a paleo-detector of age $T^n$ discussed in Sec.~\ref{sec:darkdisk}. For the subhalo scenario, on the other hand, 
\begin{equation} \label{eq:NSH}
    N_i^n({\bm \theta}; \zeta_0,\bm\zeta) = (N_0)_i^n({\bm \theta}) + M^n n_i\SH(\zeta_0, {\bm \zeta}; T^n) \;, 
\end{equation}
where $\zeta_0 = b\SH$, ${\bm \zeta} = \{m_\chi\SH, M_{\rm vir}\SH, c\SH, v_\odot\SH, T\SH\}$ and $n_i\SH$ is the subhalo signal discussed in Sec.~\ref{sec:subhalo}. Note that in Eq.~\eqref{eqn:nullspec}, the neutrino-induced background contributions scale linearly with $\Phi^{\nu_j}$, the radiogenic contributions scale linearly with $\mathcal{C}^n$, and the Milky Way halo contribution scales linearly with $\SIDDMW$. Likewise, the dark disk contribution in Eq.~\eqref{eq:Ndisk} scales linearly with $\zeta_0$ (although the subhalo contribution in Eq.~\eqref{eq:NSH} does not). 

To compute the discrimination reach, we use the maximum likelihood ratio test statistic~\cite{2012PhRvD..85c5006B}:
\begin{equation} \label{eqn:MLR}
    q(\zeta_0) = -2 \ln \left[ \frac{\LL\left( {\bm \DD} \middle| \hat{\hat{{\bm \theta}}}; \zeta_0, {\bm \zeta} \right) }{\LL\left( {\bm \DD} \middle| \hat{\bm \theta}; \hat{\zeta}_0, {\bm \zeta} \right) } \right]\;.
\end{equation}
In the numerator, $\hat{\hat{\bm \theta}}$ is the set of nuisance parameters which maximizes the likelihood for fixed values of the parameter controlling the normalization of the dark disk/subhalo signal, $\zeta_0$, and ${\bm \zeta}$. In the denominator, on the other hand, $\hat{\bm \theta}$ and $\hat{\zeta}_0$ denote the values of ${\bm \theta}$ and $\zeta_0$ which jointly maximize the likelihood $\LL$. 

We use {\it Asimov} data sets,
\begin{equation}
    {\bm \DD} = {\bm \DD}(\overline{\bm \theta}; \zeta_0^*, {\bm \zeta}) = {\bm N}(\overline{\bm \theta}; \zeta_0^*, {\bm \zeta}) \;,
\end{equation}
where ${\bm N}$ is given by Eq.~\eqref{eq:Ndisk} for the dark disk and Eq.~\eqref{eq:NSH} for the subhalo case, $\overline{\bm \theta}$ is a fiducial set of values for the nuisance parameters, $\zeta_0^*$ is the value of $\zeta_0$ for which we compute the Asimov data, and ${\bm \zeta}$ are the remaining parameters controlling the dark disk/subhalo spectrum. For the dark disk scenario, we will assume that the particles making up the smooth MW halo do not give rise to a measurable signal in paleo-detectors. Accordingly, we generate Asimov data sets for $\SIDDMW = 0$.\footnote{Note that this choice implies that the maximum likelihood estimator $\hat{\bm \theta}$ under the alternative hypothesis will lie on the boundary $\SIDDMW=0$ of parameter space in the dark disk scenario.  The assumptions of Wilks' theorem however only require that the maximum likelihood estimator $\hat{\hat{\bm \theta}}$ under the \emph{null hypothesis} is far from the boundary of parameter space.  We have verified explicitly that this assumption still holds in our analysis.} For the subhalo case, on the other hand, we will assume that the same DM particles make up the MW-halo and the subhalo, and use Asimov data sets with $\SIDDMW = \SIDDSH$ and $m_\chi\MW = m_\chi\SH$, where $\{\SIDDSH, m_\chi\SH\} \in {\bm \zeta}$.

The discrimination reach is then obtained for the different signals by computing $q(\zeta_0 = \SIDDdisk \Sigma\disk = 0)$ for the dark disk and $q(\zeta_0 = b\SH = \infty)$ for the subhalo as a function of the value $\zeta_0^*$ for which the Asimov data is generated. Values of $\zeta_0^*$ for which $q(0) \geq q_{\rm crit} = 3.84$\footnote{We are performing a nested hypothesis test for which, by Wilks' theorem~\cite{Wilks:1938dza}, the maximum log-likelihood ratio is asymptotically $\chi^2$ distributed. For a one-dimensional $\chi^2$-distribution, $\chi^2 = 3.84$ corresponds to a $p$-value of 0.05.} correspond to an ability to discriminate the time-varying component at more than 95\,\% significance. 

We will discuss our choices for $\overline{\bm \theta}$ and ${\bm \zeta}$ for the dark disk and the subhalo scenarios in Sec.~\ref{sec:results} where we present results for the discrimination reach. The code used in this work is available at: \href{\linkPaSpec}{\texttt{paleoSpec}}~\cite{PaleoSpec} for the computation of the signal and background spectra, and \href{\linkPaSens}{\texttt{paleoSens}}~\cite{PaleoSens} for the sensitivity forecasts.

\subsection{Systematic Modeling Uncertainty} \label{sec:uncertainty_stats}

In Eq.~\eqref{eqn:likelihood}, we used a Poisson likelihood to evaluate the sensitivity of a series of paleo-detectors. For samples in which the number of tracks is small, the Poisson error is relatively large. On the other hand, when a sample contains a large number of tracks, the Poisson error becomes small and \textit{systematic uncertainties} in the modeling of the spectra must be taken into account. For instance, it is not expected that our background model predictions will agree with the observed data to arbitrary precision, leading to a theoretical systematic uncertainty. This is particularly problematic when using the Asimov data set, which can be \textit{exactly} fit by the likelihood.

There are a variety of ways to account for systematic errors. Here we take a simple approach and replace the Poisson contribution to the log-likelihood in Eq.~\eqref{eqn:likelihood} with a Gaussian:
\begin{equation} \label{eq:LLGauss}
    \left[ \DD_i^n\ln N_i^n - N_i^n \right] \rightarrow -\frac{1}{2} \frac{\left(\DD_i^n - N_i^n \right)^2}{(\sigma^n_i)^2} \;,
\end{equation}
where the variance in the $i$-th bin of the $n$-th sample, $(\sigma_i^n)^2$, is given by
\begin{equation} \label{eq:Gausserr}
    (\sigma_i^n)^2 = \DD_i^n + (\sigma_{\mathrm{unc}}\, \mathcal{B}_i^n)^2\;.
\end{equation}
Here, $\mathcal{B}_i^n$ is the number of events we expect from the neutrino-induced and radiogenic backgrounds (in the $i$-th bin of the $n$-th sample), and $\sigma_{\rm unc}$ is the relative error we assign to the prediction of $\mathcal{B}_i^n$. For $\sigma_{\rm unc} = 0$, the variance $(\sigma_i^n)^2$ is just the Poisson error of the data. Hence, for $\DD_i^n \gg 1$, the sensitivity obtained after making the replacement shown in Eq.~\eqref{eq:LLGauss} will be identical to those discussed in Sec.~\ref{sec:stat} (with results in Sec.~\ref{sec:results}). Setting $\sigma_{\rm unc}$ to values larger than $\sigma_{\rm unc} = 0$, on the other hand, allows us to include systematic modeling uncertainties in the {\it shape and time dependence} of the background spectra. For example, $\sigma_{\rm unc} = 0.1$ corresponds to a 10\,\% bin-to-bin modeling uncertainty in the background spectra. In Sec.~\ref{sec:uncertainty}, we explore how our sensitivity estimates react to changes in $\sigma_{\rm unc}$ in order to demonstrate the robustness of our results.

\section{Results} \label{sec:results}

In this section, we show the discrimination reach of paleo-detectors to time-varying DM signals.\footnote{Recall that we define the discrimination reach as the smallest normalization of the DM substructure-induced signal that would allow one to discriminate such a time-varying signal from the constant signal induced by the smooth MW halo, not between a time-varying DM signal and no DM signal at all.} In Sec.~\ref{sec:DarkDiskresults} we discuss results for the case of periodic crossings through a dark disk, and in Sec.~\ref{sec:Subhaloresults}, we discuss the case of a transit through the dense central region of a subhalo. For both scenarios, we will first discuss the sensitivity under a set of fiducial assumptions on the experimental setup and then vary these assumptions to assess the robustness of our results. In Sec.~\ref{sec:uncertainty}, we explore the sensitivity of paleo-detectors in the presence of systematic modeling uncertainties (as discussed in Sec.~\ref{sec:uncertainty_stats}) for both substructure scenarios. Taken together, these results demonstrate a key finding of this paper: paleo-detectors have sensitivity to DM-substructure-induced signals for a wide variety of experimental realizations, even in the presence of significant uncertainties on nuisance parameters and systematic modeling uncertainties. 

\subsection{Dark Disk} \label{sec:DarkDiskresults}

In order to explore the sensitivity of paleo-detectors to a signal induced by a dark disk, we consider the following fiducial experimental scenario:
\begin{itemize} \itemsep0em 
    \item Number of samples: 5,
    \item High-resolution readout (sample mass $M_s^n = 10\,$mg, track length resolution $\sigma_{x_T} = 1\,$nm),
    \item Sample ages: $T^n = \{20, 40, 60, 80, 100\}$\,Myr,
    \item $^{238}$U concentration: $\mathcal{C}^n = 10^{-11}$\,g/g.
\end{itemize}
Together with $\SIDDMW = 0$, these values define $\overline{\bm \theta}$ using the notation from Sec.~\ref{sec:stat}. For the dark disk case, ${\bm \zeta} = \{m_{\chi}\disk\}$; we therefore compute the discrimination reach on a grid of fixed $m_{\chi}\disk$.

\begin{figure}
    \includegraphics[width=\linewidth]{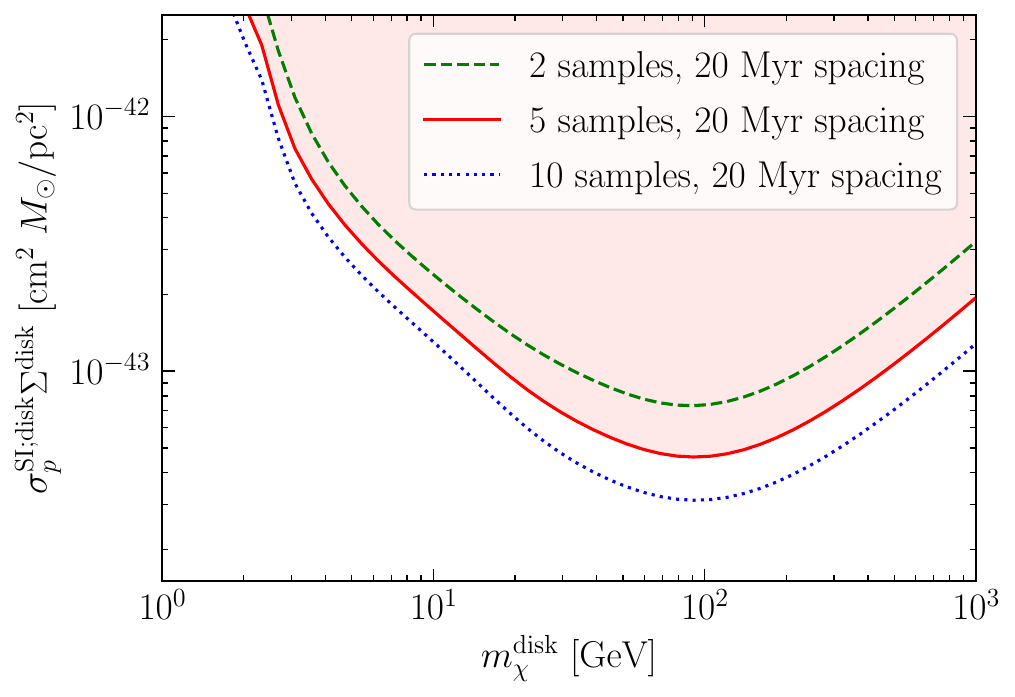}
    \caption{Here, we show the discrimination reach on the product $\SIDDdisk \Sigma\disk$ for our fiducial scenario (5 samples, solid red) along with results for 2 samples (with $T^n = \{20, 40\}\,$Myr; dotted green) and 10 samples (with $T^n = \{20, 40, \ldots, 200\}\,$Myr; dotted blue). The red shaded region indicates where the fiducial scenario has sensitivity. We use $M_s^n = 10\,$mg throughout, hence it is unsurprising that 10 samples outperform 5 samples, and that 5 samples outperform 2 samples due to the accumulated exposure. Nonetheless, it is evident that while we have selected a particular choice for our fiducial parameters, the general results are not heavily influenced by the number of samples.
    \label{Fig:diffN}}
\end{figure}

\begin{figure}
    \includegraphics[width=\linewidth]{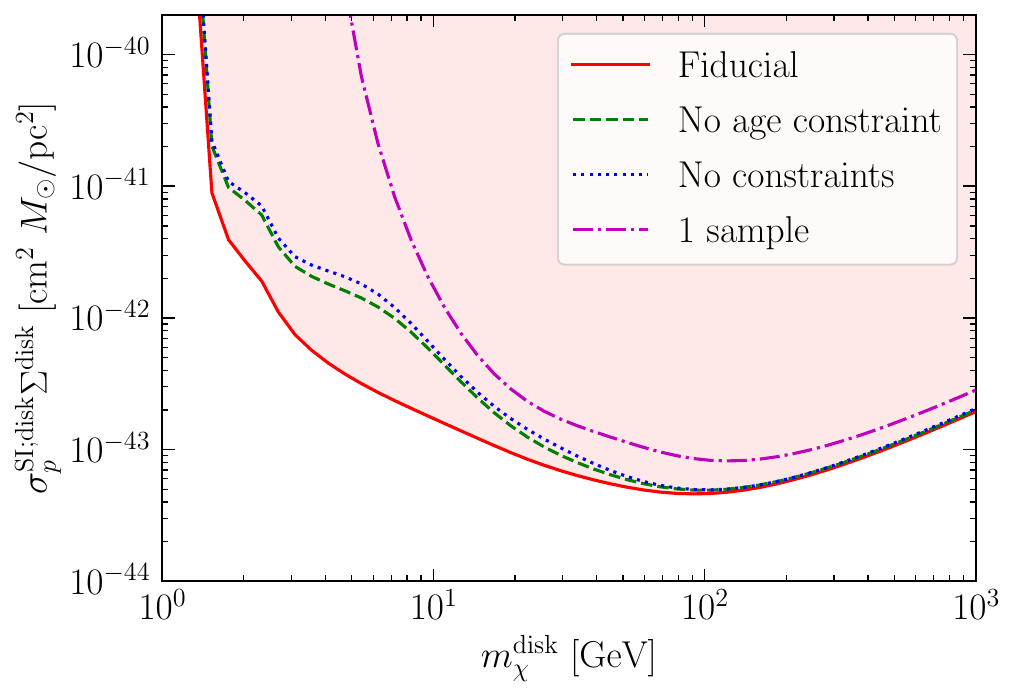}
    \caption{Here, we show that it is the use of multiple samples of differing ages that provides a large degree of the sensitivity of our fiducial results. To demonstrate this, we have compared the sensitivity of our fiducial results (solid red) to the sensitivity of a single sample, matching the total exposure (dot-dashed purple). This serves as a proxy for purely spectral discrimination. To show the power of the backgrounds themselves to provide age information even in the absence of any external age constraints, we have plotted results which correspond to the fiducial parameters, but with no external age constraint (dashed green) and with no external constraints whatsoever on sample age, sample mass, or any of the background normalizations (dotted blue). Even without any external information about these nuisance parameters, these scenarios still dramatically outperform the single-sample scenario which relies purely on differences in the shape of the track length spectra. This result indicates that the age information provided solely by the backgrounds themselves allows for strong sensitivity.
    \label{Fig:agevsspec}}
\end{figure}

In Figs.~\ref{Fig:diffN}\,--\,\ref{Fig:agevsspec}, we plot the discrimination reach on the product $\SIDDdisk \Sigma\disk$ for the fiducial scenario (shown throughout by the solid red line, with a red shaded region indicating the parameter space in which the fiducial scenario has sensitivity) together with other scenarios where we change one parameter while keeping all others fixed.\footnote{Note that here, and throughout the results section, the red shading indicates the region of the parameter space which is potentially discriminable for our fiducial scenario.} In general, these variations do not appreciably affect our sensitivity estimates. The results show that paleo-detectors can discriminate a dark disk signal from the halo for a variety of experimental realizations.

\begin{figure}
    \includegraphics[width=\linewidth]{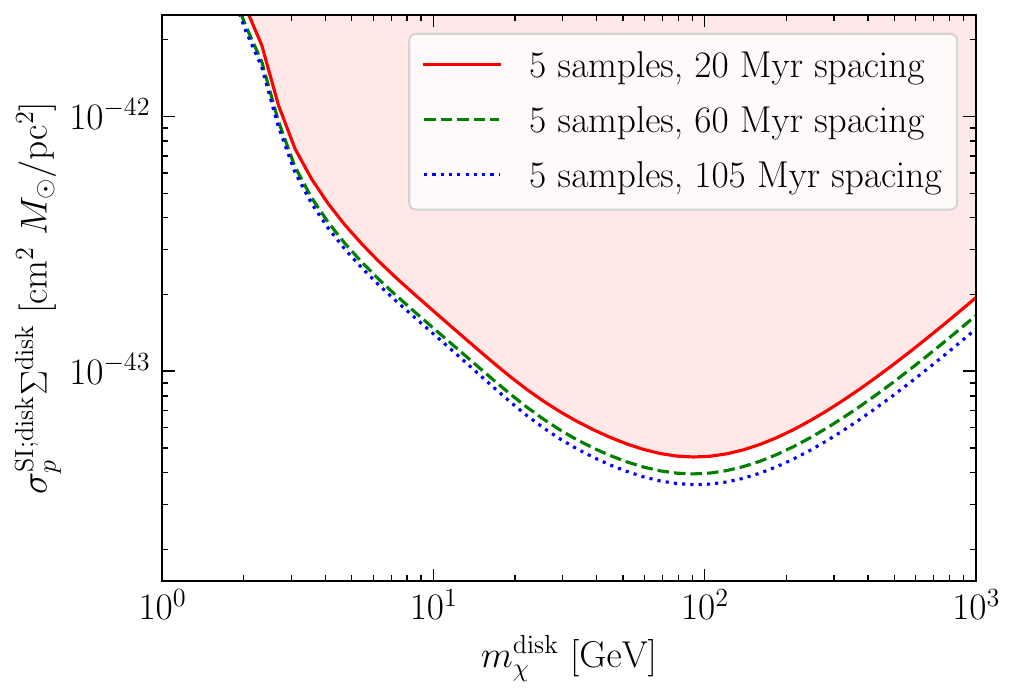}
    \caption{Same as Fig.~\ref{Fig:diffN}, but for various choices of the age spacing between samples. We show our fiducial result ($T^n - T^{n-1} = 20\,$Myr, solid red) along with results for both $T^n - T^{n-1} = 60\,$Myr spacing (dotted green) and $T^n - T^{n-1} = 105\,$Myr spacing (dotted blue) with 5 samples each. The sample masses, $M_s^n = 10\,$mg, have not been rescaled to account for differences in accumulated exposure, hence it is unsurprising that the 60\,Myr and the 105\,Myr spacing scenarios outperform the $T^n - T^{n-1} = 20\,$Myr scenario. However, it should be noted that the gain due to larger exposure saturates rapidly, hence the general results are not heavily influenced by our particular choice of age spacing.
    \label{Fig:diffspacing}}
\end{figure}

\begin{figure}
    \includegraphics[width=\linewidth]{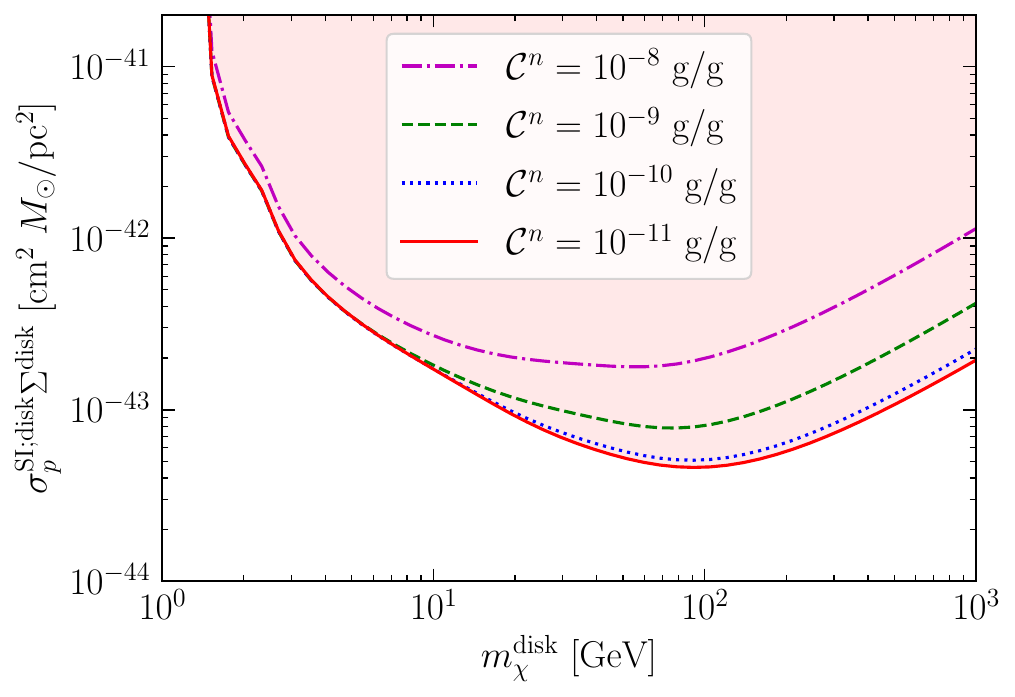}
    \caption{Same as Fig.~\ref{Fig:diffN}, but for various $^{238}$U concentrations, $\mathcal{C}^n$. We show our fiducial result ($\mathcal{C}^n = 10^{-11}\,$g/g, solid red) along with other values for $\mathcal{C}^n$ as indicated in the legend. While more radiopure samples have better sensitivity, there is little loss in sensitivity even for samples with orders of magnitude larger $\mathcal{C}^n$.
    \label{Fig:diffU}}
\end{figure}

Let us first discuss the mass dependence of the discrimination reach for our fiducial scenario. This scenario achieves the best sensitivity at $m_\chi\disk\sim 100\,$GeV, and has a qualitatively similar $m_\chi$-dependence to a conventional direct detection experiment. For masses below $\sim 100\,$GeV, the sensitivity depreciates with decreasing mass because the DM particles have lower kinetic energy and therefore give rise to softer nuclear recoils. This results in a large proportion of the track length spectrum being below the readout resolution. On the other hand, for $m_\chi\disk \gtrsim 100\,$GeV, the recoils are well-above the resolution threshold. Then, the dominant effect is that the number of signal events is proportional to the DM number density, which for fixed $\Sigma\disk$ scales with $1/m_\chi\disk$.

Now let us examine the effect of variations upon the fiducial case one by one. In Fig.~\ref{Fig:diffN}, we fix the age spacing and the sample mass, but show the sensitivity for a scenario with only two samples and a scenario with 10 samples. There is no appreciable advantage to accumulating more samples --- even just two samples have comparable sensitivity to five samples. Similarly, in Fig.~\ref{Fig:diffspacing}, we see that the sample age spacing has a negligible effect on the sensitivity. Figure~\ref{Fig:diffU} shows that changes to the $^{238}$U concentration of the samples, $\mathcal{C}^n$, make little difference at low $m_\chi\disk$, while at larger $m_\chi\disk$, cleaner samples do outperform less pure samples. This is not unexpected, as the radiogenic neutron background only begins to dominate at high masses (see Fig.~\ref{Fig:bkgs}). Note that even for $^{238}$U concentrations of $\mathcal{C}^n = 10^{-8}$\,g/g, there is less than an order of magnitude weakening of the sensitivity (in terms of $\SIDDdisk \Sigma\disk$) compared to our fiducial assumption, $\mathcal{C}^n = 10^{-11}\,$g/g. This serves as yet another indicator that paleo-detectors have strong sensitivity to the dark disk even for highly non-optimal experimental scenarios.

While the MW halo and dark disk produce different recoil spectra, this is \textit{not} what dominates the sensitivity. Rather, it is truly information about how the signal has varied over time. We show this in Fig.~\ref{Fig:agevsspec}. As before, the solid red curve is the fiducial experimental realization. When the analysis is repeated using only a single sample (dot-dashed purple) the sensitivity weakens appreciably. This is because when only one sample is used, the only means to discriminate the dark disk and MW halo is through their spectral differences. While this shows that having samples of various ages is critical to the sensitivity, external information on the sample ages need not be provided --- in fact, just the relative normalization of backgrounds in a sample provides a handle on its age. This can be seen by removing the external age constraints\footnote{Technically, removing the constraint on a nuisance parameter corresponds to omitting the corresponding entry in the sum in the second line of Eq.~\eqref{eqn:likelihood} or, equivalently, taking the corresponding uncertainty $c_j \to \infty$.} from the fiducial scenario (dashed green), which does not dramatically affect the sensitivity. Furthermore, explicit external constraints on the background normalizations are also not needed to extract age information from the backgrounds (dotted blue).\footnote{More generally, external constraints on these nuisance parameters are typically not required to maintain competitive sensitivity to the smooth MW halo, even in the single sample case (see Ref.~\cite{Baum:2021jak}).} Taken together, these results demonstrate that the discriminating power of paleo-detectors comes from the varying ages of the samples, and that external measurements of these ages need not be provided to yield strong sensitivity. It is a key result of this paper that the backgrounds themselves can provide age information even in the absence of external constraints, further bolstering the case for paleo-detectors as robust probes of our Galactic history.

Existing constraints on a dark disk arise from the kinematics of stars in the MW~\cite{Kramer:2016dqu,Kramer:2016dew,Schutz:2017tfp,Buch:2018qdr,Widmark:2018ylf,Widmark:2020vqi,Widmark:2021gqx}. However, these constraints arise from gravitational interactions and are thus insensitive to the scattering cross section of the DM comprising the dark disk. These astrometric limits are $\Sigma\disk \lesssim \mathcal{O}(5)\,\mathrm{M}_{\odot}/\text{pc}^2$ for thin disks with disk height $Z\disk \lesssim 10$\,pc. Hence, for cross sections $\SIDDdisk \gtrsim 10^{-43}\,{\rm cm}^2$, paleo-detectors could probe surface densities $\Sigma\disk$ far below current astrometric constraints. Note that unlike $\SIDDMW$, $\SIDDdisk$ is \textit{unconstrained} by conventional direct detection experiments, as we have not transited the disk in the last Myr. Paleo-detectors are sensitive to the product $\SIDDdisk\Sigma\disk$, while astrometric measurements are sensitive directly to $\Sigma\disk$. Thus, combining the results of both methods would provide information on $\SIDDdisk$, motivating further developments of both techniques.

\subsection{Subhalos} \label{sec:Subhaloresults}

In this section, we discuss the sensitivity of a series of paleo-detectors to the signal induced by the Solar System traversing the dense inner region of a subhalo during the past $\sim 1\,$Gyr. As discussed above, we assume that the DM particles making up the subhalo and the MW halo have the same DM-proton scattering cross section and mass, $\SIDDSH = \SIDDMW$ and $m_\chi\SH = m_\chi\MW$, respectively. Thus, we will assume that the Asimov data is the sum of the various backgrounds, the DM MW halo, and the subhalo signal. In addition to $\SIDDSH$ and $m_\chi\SH$, the subhalo signal is controlled by the mass ($M_{\rm vir}\SH$) and concentration parameter ($c\SH$) of the subhalo, the relative speed of the encounter ($v_\odot\SH$), the time of closest approach ($T\SH$), and the impact parameter between the Solar System and the subhalo ($b\SH$). As discussed above, we will mainly be interested in subhalos in the mass range $10^4\,M_\odot \lesssim M_{\rm vir}\SH \lesssim 10^8\,M_\odot$, and the Solar System would encounter a typical MW subhalo with $v_\odot\SH$ of a few hundred km/s. Using the notation from Sec.~\ref{sec:stat}, we have ${\bm \zeta} = \{m_\chi\SH, M_{\rm vir}\SH, c\SH, v_\odot\SH, T\SH\}$ and $\zeta_0 = b\SH$.

Rather than working with the DM particle parameters directly, as in the dark disk case, we instead consider parameters which characterize the subhalo crossing. The impact parameter $b\SH$ approximately governs the signal normalization, in analogy to $\SIDDdisk\Sigma\disk$ in the disk case. Meanwhile, $v_\odot\SH$ governs the spectral shape of the signal, in analogy with $m_\chi\disk$ in the disk case. For our fiducial scenario, we fix the remaining parameters in ${\bm \zeta}$ to some reference values and then explore how our sensitivity changes as we vary each.

We use the following fiducial parameters for the subhalo-induced signal:
\begin{itemize}
\itemsep0em 
    \item DM mass and cross section: $m_\chi\SH=500\,\text{GeV}$, $\SIDDSH = 5\times10^{-46}\,\text{cm}^2$ --- these are compatible with the current null-results from conventional direct detection experiments~\cite{Aprile:2018dbl},
    \item Subhalo parameters: $M_{\rm vir}\SH=10^6\,M_\odot$ and $c\SH=65$,\footnote{Recall that this choice was informed by the mass-concentration relation from Ref.~\cite{Moline:2016pbm}.}
    \item Time of closest approach: $T\SH=500\,\text{Myr}$,
\end{itemize}
and we will consider the fiducial experimental setup (i.e., values for $\overline{\bm \theta}$):
\begin{itemize} \itemsep0em 
    \item Number of samples: 5,
    \item Low-resolution readout (sample mass $M_s^n = 100\,$g and track length resolution $\sigma_{x_T} = 15\,$nm),
    \item Sample ages: $T^n = \{200, 400, 600, 800, 1000\}$\,Myr,
    \item $^{238}$U concentration: $\mathcal{C}^n = 10^{-11}$\,g/g.
\end{itemize}
In contrast to the dark disk scenario discussed in Sec.~\ref{sec:DarkDiskresults}, the longer subhalo-induced tracks are detectable even in the low-resolution scenario, allowing us to consider much larger sample masses. Furthermore, note that here we use much older rocks than for the dark disk scenario. While the Solar System would pass through a dark disk every $\sim 45\,$Myr, the chance of passing through a detectably dense region of a subhalo is relatively small, even within the past Gyr (see Appendix~\ref{app:subhalo}). Thus, in order to increase the chance of a subhalo encounter during the exposure time, it is beneficial to use samples with large $T^n$. 

\begin{figure}
    \includegraphics[width=\linewidth]{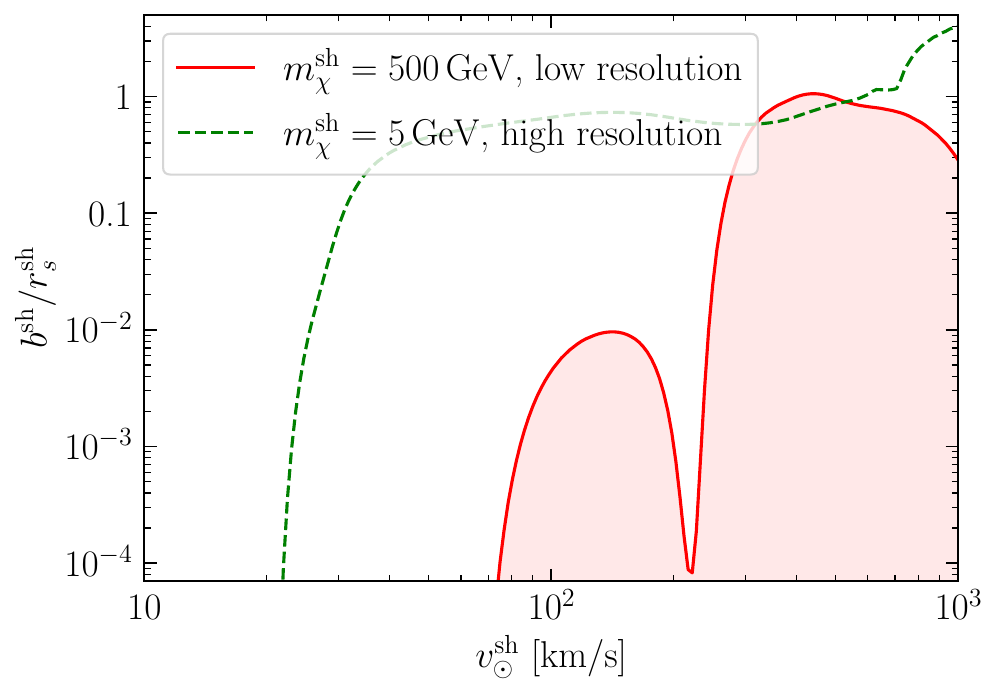} 
    \caption{Here, we plot the discrimination reach, $b\SH_{\rm max}$, as a function of the relative velocity of the subhalo, $v_\odot\SH$. Note that paleo-detectors could discriminate a subhalo-induced signal from the signal due to only the smooth MW halo for all impact parameters \emph{below} the plotted results (indicated by the red shaded region for the fiducial scenario). In solid red, we show our fiducial result, where we consider a DM mass of $m_\chi\SH = 500\,\text{GeV}$ with a DM-proton cross section $\SIDDSH = 5\times10^{-46}\,\text{cm}^2$, just below current limits from conventional direct detection experiments~\cite{Aprile:2018dbl}. Here, we assume the low-resolution readout scenario ($M_s^n = 100\,\text{g}$, $\sigma_{x_T}=15\,\text{nm}$). We also show the case of a smaller DM mass $m_\chi\SH = 5\,\text{GeV}$ with $\SIDDSH = 10^{-43}\,\text{cm}^2$ (dashed green). Due to the softer track length spectra, we use the high-resolution readout scenario ($M_s^n = 10\,$mg, $\sigma_{x_T}=1\,\text{nm}$) for the $m_\chi\SH = 5\,\text{GeV}$ benchmark.}
    \label{Fig:subhalo}
\end{figure}

\begin{figure}
    \includegraphics[width=\linewidth]{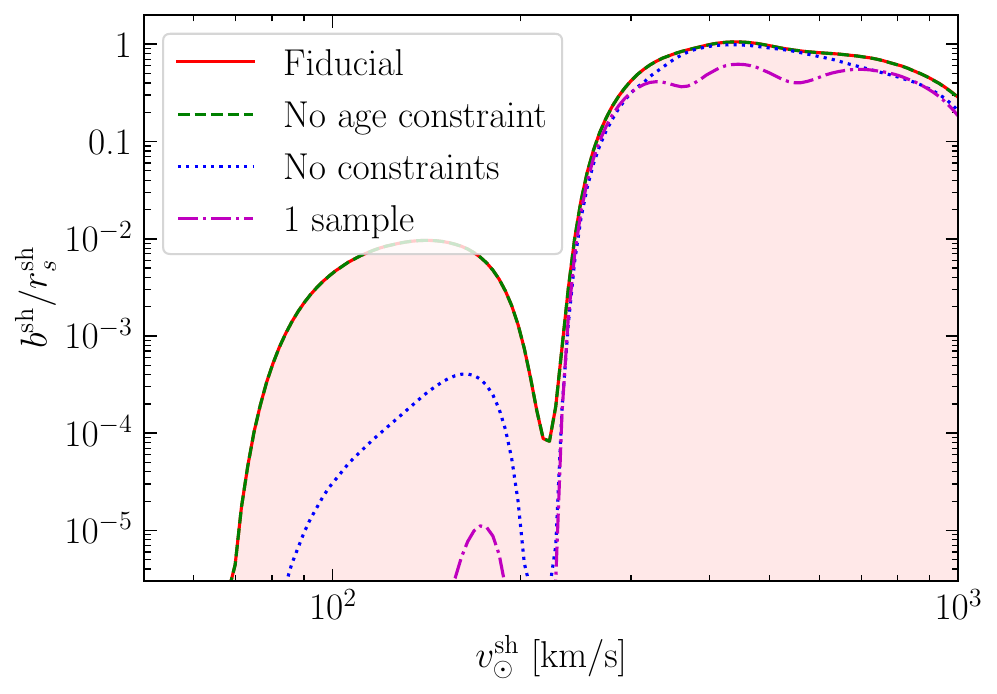}
    \caption{Same as Fig.~\ref{Fig:agevsspec}, but for the subhalo case. We compare our fiducial results (solid red) to the case of a single sample with the same total exposure (dot-dashed purple). The latter only utilizes spectral information, without relying on any temporal information. Also shown are results with no external age constraints (dashed green) and no external constraints at all (dotted blue); these demonstrate that the backgrounds themselves contain temporal information, even without external constraints. At low velocities, it is clear that a significant contribution to the sensitivity comes from temporal information. In contrast, at high velocities, spectral information alone can achieve considerable sensitivity. Furthermore, external constraints play very little role at high velocities, and nearly the same sensitivity can be reached without any external constraints on the nuisance parameters.}
    \label{Fig:subhalo_constraints}
\end{figure}

In Figs.~\ref{Fig:subhalo}\,--\,\ref{Fig:subhalo_constraints} we show results for the sensitivity of a series of paleo-detectors to a subhalo transit. As discussed above, we use the plane spanned by $v_\odot\SH$ and $b\SH$ to show our sensitivity projections; note that we plot the discrimination reach $b\SH_{\rm max}$ in units of the scale radius, $r_s\SH$, of the subhalo considered. We show the fiducial scenario described above with the solid red line in Figs.~\ref{Fig:subhalo}\,--\,\ref{Fig:subhalo_constraints}, and compare these results to the sensitivity for different assumptions on the parameters controlling the subhalo-induced signal in each figure.

\begin{figure}
    \includegraphics[width=\linewidth]{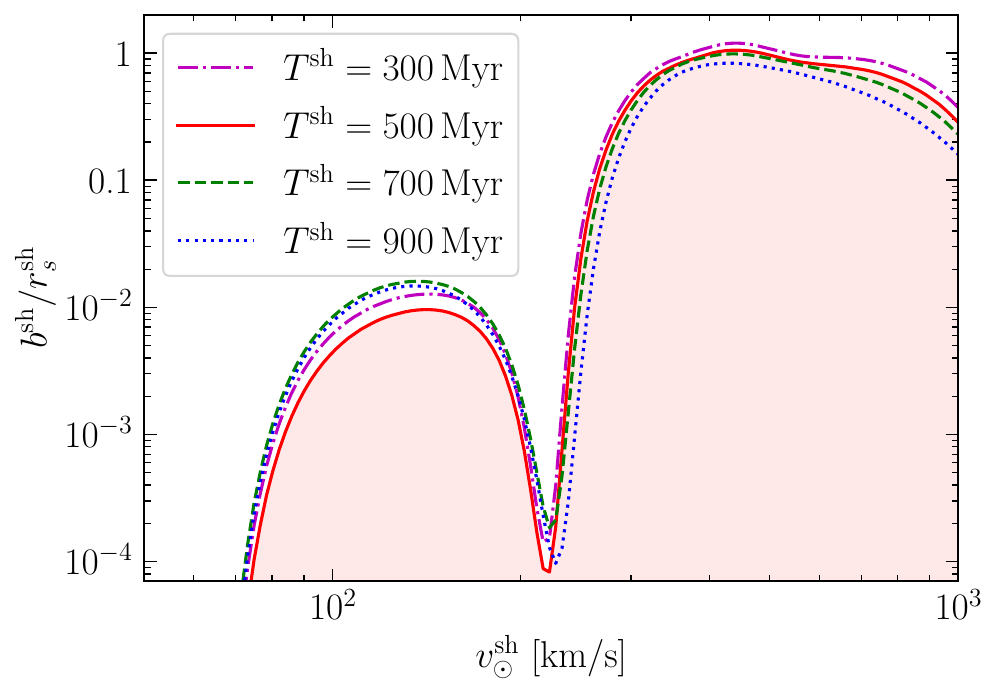}
    \caption{Here, we show the dependence of our result on the time of closest approach to the subhalo, $T\SH$. The solid red curve shows our fiducial assumption of $T\SH=500\,$Myr, while the other curves are for values of $T\SH$ as denoted in the legend. Note that these values of $T\SH$ are chosen such that a different number of samples have ages $T^n > T\SH$ in each case.}
    \label{Fig:closest_approach}
\end{figure}

\begin{figure}
    \includegraphics[width=\linewidth]{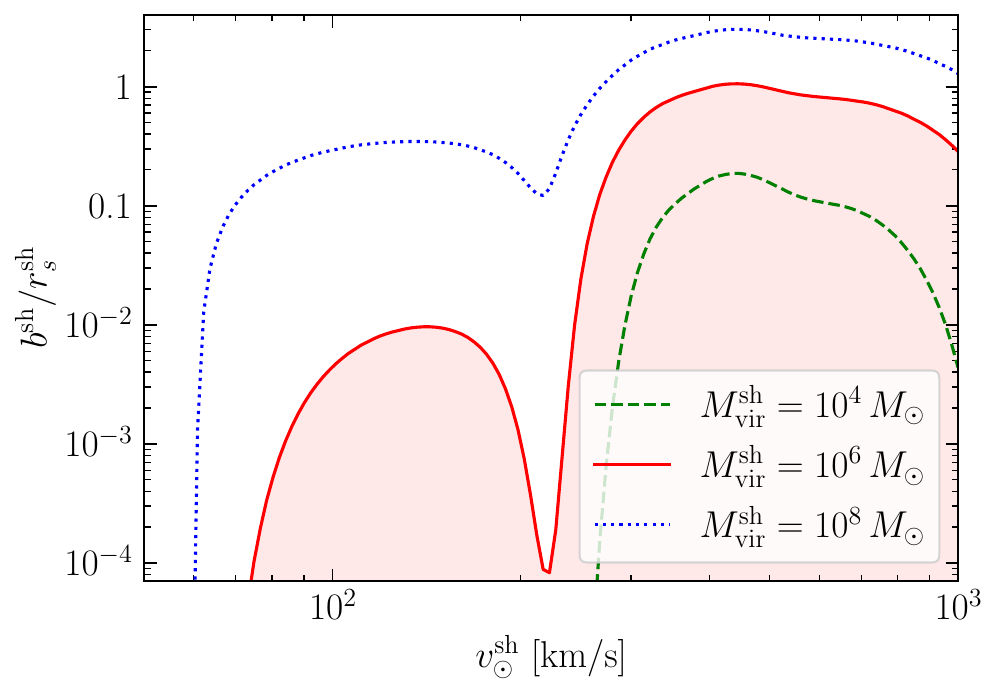}
    \caption{Here, we show the effect of varying the subhalo mass, $M_{\rm vir}\SH$. We show our fiducial case ($M_{\rm vir}\SH = 10^6\,M_\odot$, solid red), along with a less massive ($M_{\rm vir}\SH = 10^4\,M_\odot$, dashed green) and more massive ($M_{\rm vir}\SH = 10^8\,M_\odot$, dotted blue) subhalo. Note that we normalize $b\SH$ by the scale radius of the subhalo, $r_s\SH$, which varies as the subhalo mass is changed. We see that the effect of changing $M_{\rm vir}\SH$ on the sensitivity parametrized by $b\SH_{\rm max}(v_\odot\SH)/r_s\SH$ is much larger at encounter speeds $v_\odot\SH \lesssim 200\,$km/s than at larger $v_\odot\SH$.}
    \label{Fig:subhalo_masses}
\end{figure}

Let us begin the discussion of our results with the fiducial case. The features of this sensitivity curve can be understood by considering the relevant backgrounds. At low $v_\odot\SH$, where the subhalo track length spectrum is softer, the dominant background is the single-$\alpha$ background (which is broadened by the low-resolution readout; see left panel of Fig.~\ref{Fig:bkgs}). However, at larger $v_\odot\SH$, where the subhalo spectrum is harder, the radiogenic neutrons become the dominant background source. These two regimes are separated by a characteristic ``dip'' in sensitivity at around $200\,\mathrm{km/s}$, where the subhalo-induced track length spectrum mimics the shape of the single-$\alpha$ background spectrum. Our sensitivity is weaker at velocities below this dip than above it due to the different normalization of the backgrounds at short and long track lengths, see the left panel of Fig.~\ref{Fig:bkgs}.

In Fig.~\ref{Fig:subhalo}, we also show results for a lower DM mass, $m_\chi\SH = 5\,\text{GeV}$ (with DM-proton cross section of $\SIDDSH = 10^{-43}\,\text{cm}^2$, compatible with current upper limits~\cite{Aprile:2019xxb}) in dashed green. Because the track length spectra for the $m_\chi\SH = 5\,$GeV are softer than for $m_\chi\SH = 500\,$GeV, we use the high-resolution readout for this alternative scenario. Note that we do not observe the same drastic ``dip'' in sensitivity as for our fiducial case; due to the higher track-length resolution, the single-$\alpha$ background spectrum is not as broad and thus has less effect on the sensitivity (see Fig.~\ref{Fig:bkgs}). The exquisite track-length resolution of the high-resolution scenario also provides sensitivity to relatively small values of $v_\odot\SH$, where the induced track-length spectra are short.

\begin{figure}
    \includegraphics[width=\linewidth]{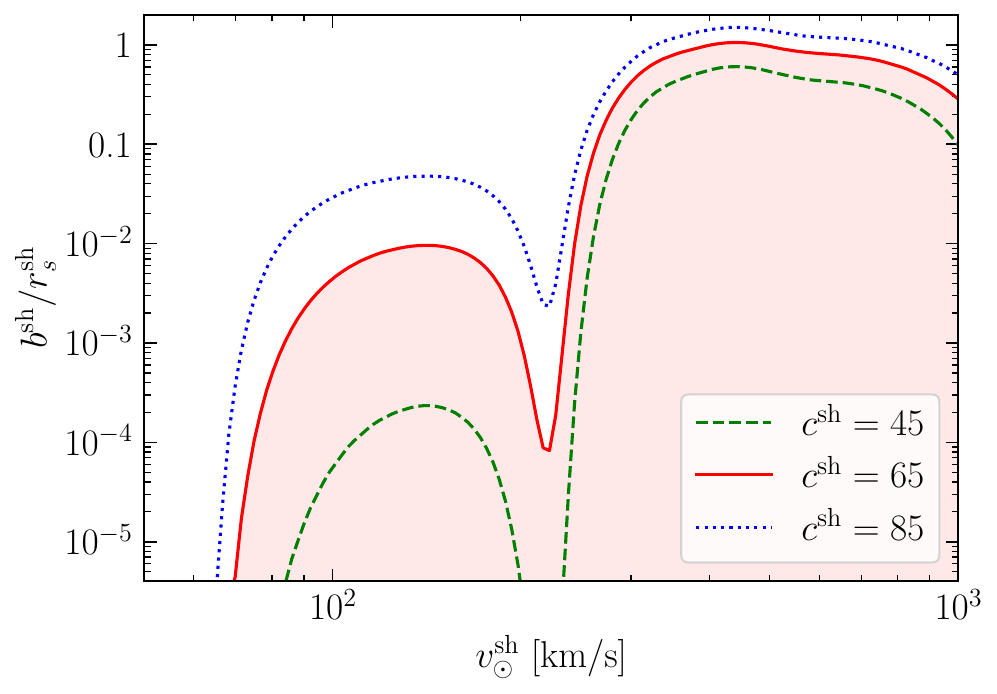}
    \caption{Here, we show the effect of varying the concentration parameter of the subhalo, $c\SH$, on the sensitivity to a subhalo signal. Our fiducial value ($c\SH = 65$) is shown in solid red, along with $c\SH = 45$ (dashed green) and $c\SH = 85$ (dotted blue). Note that the vertical axis is normalized by the scale radius, which also varies with the concentration parameter. At low relative velocities $v_\odot\SH \lesssim 200\,$km/s, the concentration parameter has a sizeable effect on the sensitivity, while at higher $v_\odot\SH$, the effect is small.}
    \label{Fig:concentration_params}
\end{figure}

\begin{figure}
    \includegraphics[width=\linewidth]{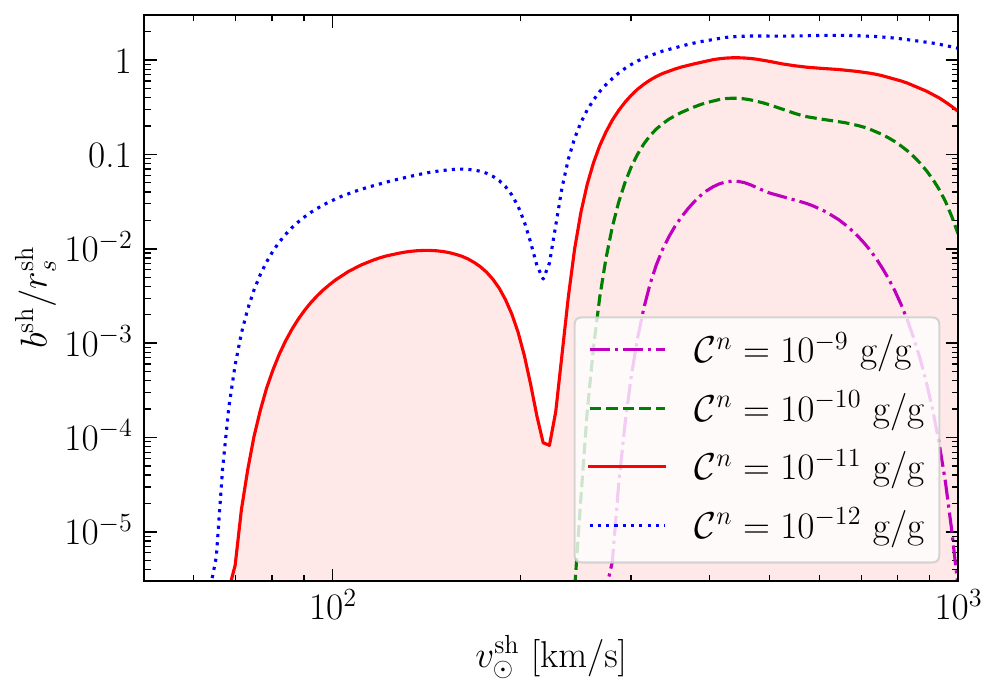}
    \caption{Here, we show how the $^{238}$U concentration of our samples $\mathcal{C}^n$, affects our subhalo result. Our fiducial case ($\mathcal{C}^n = 10^{-11}\,$g/g) is shown in solid red, and the remaining curves show the sensitivity for both larger and smaller values of $\mathcal{C}^n$ as labeled in the legend. At low encounter speeds, $v_\odot\SH \lesssim 200\,$km/s, a uranium concentration of $\mathcal{C}^n \lesssim 10^{-11}\,$g/g is necessary to achieve any appreciable sensitivity, while at higher velocities, paleo-detectors could still be sensitive to a subhalo encounter even for $\mathcal{C}^n \gg 10^{-11}\,$g/g.}
    \label{Fig:subhalo_Uconc}
\end{figure}

\begin{figure*}
    \includegraphics[width=0.49\linewidth]{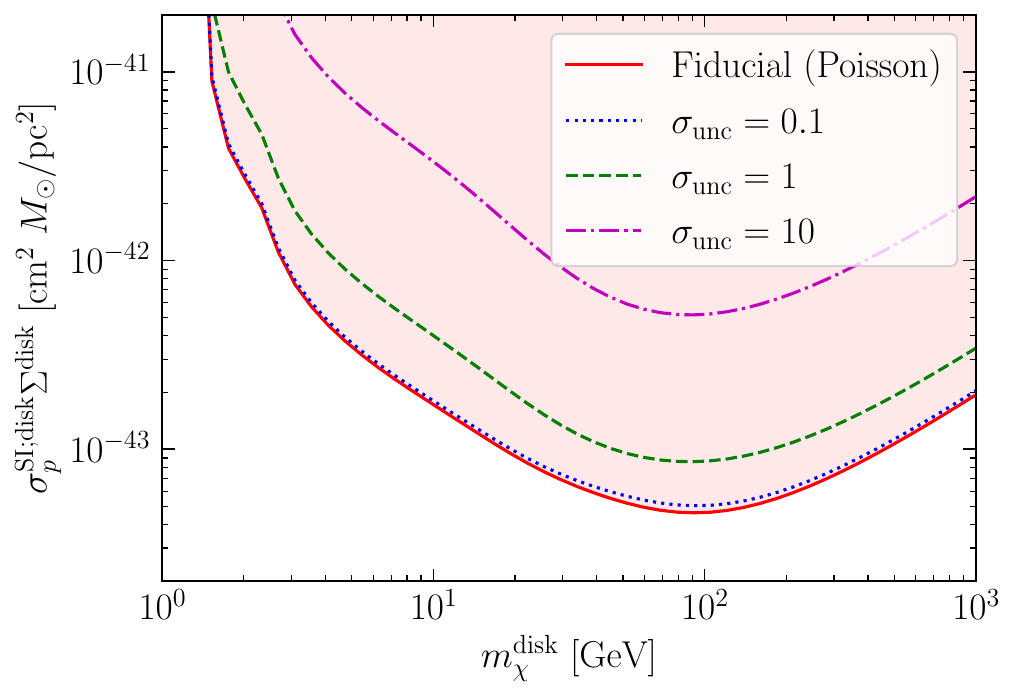}
    \includegraphics[width=0.49\linewidth]{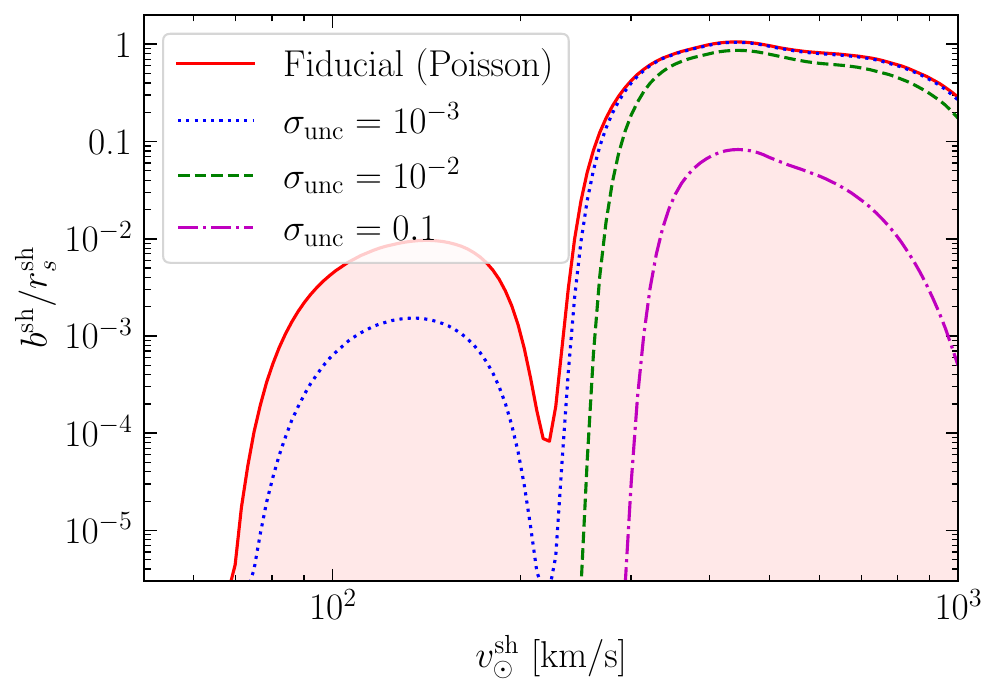}
    \caption{{\it Left:} Discrimination reach on $\SIDDdisk \Sigma\disk$ for our fiducial assumptions for the dark disk scenario (see Sec.~\ref{sec:DarkDiskresults}) using the Poisson likelihood as in Sec.~\ref{sec:DarkDiskresults} (solid red), and when using the Gaussian likelihood to account for $\sigma_{\rm unc} = 0.1$ (dotted blue), $\sigma_{\rm unc} = 1$ (dashed green), and $\sigma_{\rm unc} = 10$ (dot-dashed purple) bin-to-bin modeling uncertainties on the shape of the background spectra. {\it Right:} Discrimination reach $b\SH_{\rm max}$ for the subhalo scenario using the same fiducial assumptions as in Sec.~\ref{sec:Subhaloresults}. Here, the different lines are for $\sigma_{\rm unc} = 10^{-3}$ (dotted blue), $\sigma_{\rm unc} = 10^{-2}$ (dashed green), and $\sigma_{\rm unc} = 0.1$ (dash-dotted purple).}
    \label{Fig:diffgauss}
\end{figure*}

In Fig.~\ref{Fig:closest_approach}, we vary $T\SH$, the time of closest approach, demonstrating that $b\SH_{\rm max}$ is virtually independent of $T\SH$ as long as at least one sample is old enough to have recorded the subhalo signal. In Figs.~\ref{Fig:subhalo_masses}\,--\,\ref{Fig:subhalo_Uconc}, we vary the subhalo mass ($M_{\rm vir}\SH$), its concentration parameter ($c\SH$), and the $^{238}$U concentration in the samples ($\mathcal{C}^n$). Unlike $T\SH$, we see that changes to these parameters do have an appreciable impact on the sensitivity. The dependence of the sensitivity on $M_{\rm vir}\SH$ and $c\SH$ (see Figs.~\ref{Fig:subhalo_masses} and~\ref{Fig:concentration_params}, respectively) is straightforward to understand: the larger $M_{\rm vir}\SH$ and $c\SH$ are, the longer the Solar System (and the paleo-detectors on Earth) will spend in the dense region of the subhalo for the same encounter speed, $v_\odot\SH$, even when the ratio $b\SH/r_s\SH$ stays the same (note that $r_s\SH$ grows with both $M_{\rm vir}\SH$ and $c\SH$).\footnote{The effect changing $M_{\rm vir}\SH$, $c\SH$, and $\mathcal{C}^n$ has on the sensitivity is much smaller at $v_\odot\SH \gtrsim 200\,$km/s than at lower $v_\odot\SH$. This is because the normalization of subhalo-induced signal has a steeper dependence on $b\SH$ at larger $b\SH$, and we find larger $b\SH_{\rm max}$ at larger $v_\odot\SH$; the line-of-sight integral in Eq.~\eqref{eq:dndE_SH} scales as $\log b\SH$ for $b\SH \ll r_s\SH$ and as $(b\SH)^{-2}$ for $b\SH \gg r_s\SH$. Thus, a smaller change in $b\SH$ is required to compensate for reduced signal/increased backgrounds when $b\SH$ is higher.} The dependence of the sensitivity on $\mathcal{C}^n$ (see Fig.~\ref{Fig:subhalo_Uconc}) is much stronger than in the dark disk scenario (see Fig.~\ref{Fig:diffU}). This is because, for our fiducial scenario, radiogenics are the dominant background for the subhalo signal (see Figs.~\ref{Fig:bkgs} and~\ref{Fig:sigs}). Recall that we expect most subhalo encounters to occur in the $v_\odot\SH \gtrsim 200\,$km/s regime, where our results are less sensitive to $M_{\rm vir}\SH$, $c\SH$, and $\mathcal{C}^n$.

Finally, Fig.~\ref{Fig:subhalo_constraints} compares the importance of spectral and temporal information for our sensitivity (analogous to Fig.~\ref{Fig:agevsspec} in the dark disk case). This figure shows our fiducial result (solid red), compared with the result for a single sample (dot-dashed purple) with $M_s^n$ adjusted so that the total exposure matches the fiducial case. The single sample case serves as a proxy for purely spectral discrimination, as a single sample at a particular age cannot provide information about the temporal dependence of a signal. We see that at high velocities, spectral information alone is sufficient to discern the signal. However, at low velocities, temporal information is necessary in order to achieve significant sensitivity. Figure~\ref{Fig:subhalo_constraints} also shows the result when the age constraint is removed (dashed green) and the result when all external constraints are removed (dotted blue). These demonstrate the crucial point that external information about the age of the sample is not necessary to achieve sensitivity. Furthermore, even when {\it all} external constraints on the nuisance parameters are removed, the sensitivity at $v_\odot\SH \gtrsim 200\,$km/s is essentially unchanged.

\subsection{Systematic Modeling Uncertainty} \label{sec:uncertainty}

In this section we discuss the discrimination reach of \textit{both} the dark disk and subhalo scenarios in the presence of systematic modeling uncertainties using the method laid out in Sec.~\ref{sec:uncertainty_stats}. In Fig.~\ref{Fig:diffgauss} we show the discrimination reach for both the dark disk (left panel) and the subhalo (right panel) scenarios for a range of values of $\sigma_{\rm unc}$. In both panels, the solid red line shows the fiducial results from Secs.~\ref{sec:DarkDiskresults}/\ref{sec:Subhaloresults}, using the Poisson likelihood as discussed in Sec.~\ref{sec:stat}. The remaining curves show results after replacing the Poisson contribution in Eq.~\eqref{eqn:likelihood} with a Gaussian likelihood, Eq.~\eqref{eq:LLGauss}, for a range of values for the bin-to-bin background modeling uncertainty, $\sigma_{\rm unc}$. For context, a value of $\sigma_{\rm unc} = 0.1$ represents an allowed bin-to-bin variation in the number of events of 10\%. A value of $\sigma_{\rm unc} = 1-10$ therefore practically means we have little to no information about the shape of the background, whereas $\sigma_{\rm unc} = 10^{-3}$ means that the shapes of each of the various background contributions are constrained very well.

Let us first discuss the dark disk scenario shown in the left panel of Fig.~\ref{Fig:diffgauss}. For values of $\sigma_{\rm unc} = 0.1$, we find virtually no difference in the sensitivity compared to our fiducial analysis. Even including a large bin-to-bin modeling uncertainty of $\sigma_{\rm unc} = 1$, we find sensitivity only a factor of $\lesssim 2$ worse than for the fiducial case. For the rather extreme assumption of $\sigma_{\rm unc} = 10$, the sensitivity depreciates by about an order of magnitude in terms of the smallest $\SIDDdisk \Sigma\disk$ that could be probed. These results underline that our sensitivity forecasts for the dark disk scenario are very robust to mismodeled background spectra.

In the right panel of Fig.~\ref{Fig:diffgauss} we show the effect of including background modeling uncertainties for the subhalo scenario. In this case, the modeling uncertainty has a much greater effect, and we show results for smaller values of the bin-to-bin modeling uncertainty, $\sigma_{\rm unc} = \{10^{-3}, 10^{-2}, 0.1\}$. Recall that in contrast to the dark disk scenario, for which we were forced to use the high-resolution readout scenario due to the soft signal spectrum, in the subhalo case, the long track lengths allowed us to use the low-resolution scenario. Due to the much larger exposure of the low-resolution readout scenario, the number of background events is larger. Thus, statistical uncertainties will be smaller and systematic uncertainties will play a greater role (see also Fig.~\ref{Fig:bkgs}). For low speeds of the subhalo relative to the Solar System, $v_\odot\SH \lesssim 200\,$km/s, background modeling uncertainties as small as $\sigma_{\mathrm{unc}}=10^{-2}$ can drastically reduce the sensitivity. For higher $v_\odot\SH$, however, the effect is less dramatic and we find an appreciable loss of sensitivity only for $\sigma_{\mathrm{unc}} \gtrsim 0.1$; recall that we would expect most subhalo encounters to occur with $v_\odot\SH \gtrsim 200\,$km/s. We also note that the dominant backgrounds in the subhalo analysis are radiogenic. The shape of the radiogenic backgrounds are relatively straightforward to calibrate experimentally, for example, by measuring the track length spectrum (in the same mineral and using the same readout technique as for the search) in samples with larger $^{238}$U concentrations. Thus, we expect the background modeling uncertainties for the subhalo search to be relatively small, making our forecasts in Sec.~\ref{sec:Subhaloresults} robust.

Finally, we note that throughout this paper, we have assumed that the background components are \textit{not} intrinsically varying on time-scales $\gtrsim \mathcal{O}(1)\,$Myr relevant for paleo-detectors.\footnote{Note that variations on time-scales short compared to the age of any paleo-detector sample, e.g. annual modulations of the solar neutrino background from the ellipticity of Earth's orbit around the Sun, would average out over the relevant timescales.} If, for instance, the MW halo signal is for some reason larger in the Galactic disk, a paleo-detector may not be able to clearly distinguish this from the signal induced by a dark disk without additional information from other experimental probes. We leave a detailed exploration of the effects of time-varying backgrounds on the sensitivity for future work.

\section{Conclusion} \label{sec:conclusion}

In this paper, we have shown that paleo-detectors have a unique ability to measure the temporal dependence of signals over Myr to Gyr timescales. We have chosen two representative examples to showcase this ability: first, the signal induced by periodic transits through a dark disk, and second, the signal from a single passage through a dark matter (DM) subhalo. In both cases, we have shown that paleo-detectors could discriminate such time-varying signals from the uniform Milky Way (MW) halo signal for a wide variety of experimental realizations (e.g. number of samples, sample ages, radiopurity of the samples). More specifically, in the case of a dark disk, we have shown that reading out the tracks in as few as two samples could allow one to probe surface densities well below those probed by astrometric measurements of stars, $\Sigma\disk < \mathcal{O}(5)\,M_\odot/{\rm pc}^2$~\cite{Kramer:2016dqu,Schutz:2017tfp,Widmark:2018ylf,Buch:2018qdr,Widmark:2021gqx}, if the DM-proton cross section of the DM making up the dark disk is $\SIDDdisk \gtrsim 10^{-43}\,{\rm cm}^2$. For a subhalo transit, we showed that for subhalo encounters with relative speeds $v\SH_{\odot} \gtrsim \mathcal{O}(200)\,\mathrm{km/s}$ with respect to the Solar System and impact parameters $b\SH/r_s\SH \lesssim \mathcal{O}(1)$, a series of paleo-detectors could probe subhalo masses in a currently unconstrained part of the (sub)halo mass function. In standard cosmology, such subhalo encounters are rare. Hence, observing a signal from a subhalo in paleo-detectors would provide evidence for an enhanced subhalo mass function as could e.g. arise from nonstandard cosmology.

In many contexts, no independent measurement of the age of the samples is necessary to perform this discrimination --- the relative normalizations of the backgrounds in the different samples themselves can provide the requisite timing information. Together with the results of Sec.~\ref{sec:uncertainty}, where we demonstrated sensitivity even under large systematic modeling uncertainties in the backgrounds, this indicates that paleo-detectors are robust and flexible probes of time-varying signals which could provide invaluable information about the structure of our Galaxy.

Although we have focused on two specific scenarios, the results are general and the framework developed here can be easily applied to other time varying signals. Previous work has already demonstrated the power of paleo-detectors to measuring changes of the neutrino fluxes from the Sun~\cite{Arellano:2021jul}, Galactic supernovae~\cite{Baum:2019fqm}, or cosmic rays interacting with the atmosphere of the Earth~\cite{Jordan:2020gxx}. In order to facilitate further work, we provide ready-to-use codes which can be easily adapted to new signals: \href{\linkPaSpec}{\texttt{paleoSpec}}~\cite{PaleoSpec} for the computation of track-length spectra, and \href{\linkPaSens}{\texttt{paleoSens}}~\cite{PaleoSens} for the sensitivity forecasts. 
The broad sensitivity we have shown across many experimental realizations further motivates experimental work towards realizing paleo-detectors --- the history of the Galaxy may be revealed in a simple handful of rocks.

\begin{acknowledgements}
The authors would like to thank Katie Freese, Peter Graham, and Patrick Stengel for invaluable discussions. 
SB, SK, and WD acknowledge support by NSF Grant PHY-2014215, DOE HEP QuantISED award \#100495, and the Gordon and Betty Moore Foundation Grant GBMF7946.
SK also acknowledges support by NSF Grant DGE-1656518.
TE acknowledges support by the Vetenskapsr{\aa}det (Swedish Research Council) through contract No. 638-2013-8993 and the Oskar Klein Centre for Cosmoparticle Physics. Some of the computing for this project was performed on the Sherlock cluster. We would like to thank Stanford University and the Stanford Research Computing Center for providing computational resources and support that contributed to these research results. 
We acknowledge the use of the Python scientific computing packages NumPy~\cite{numpy,Harris:2020xlr} and SciPy~\cite{Virtanen:2019joe}, as well as the graphics environment Matplotlib~\cite{Hunter:2007}. 
\end{acknowledgements}

\begin{appendix}

\section{Subhalo transit rate}
\label{app:subhalo}

In this Appendix, we compute the probability for a detectable subhalo encounter for a given population of subhalos, characterized by the subhalo mass function, $\dd N\SH / \dd M_{\rm vir}\SH$, the mass-concentration relation, $c\SH(M_{\rm vir}\SH)$, the spatial distribution of subhalos in the MW, and the speed distribution of the subhalos relative to the Solar System, $f(v_\odot\SH)$.

Let us denote the number density of subhalos at the location of the Solar System with $n\SH$. The number of subhalos which have come within a distance $\leq b\SH_{\rm max}$ during the time $\Delta T$ is then
\begin{equation}
    n\SH \pi (b\SH_{\rm max})^2 v_\odot\SH \Delta T \;,
\end{equation}
where $v_\odot\SH$ is the speed of the subhalo relative to the Solar System. Since $b\SH_{\rm max} = b\SH_{\rm max}(M\SH_{\rm vir}, c\SH, v_\odot\SH)$, we must average over these parameters. To this end, we normalize the subhalo mass function into a probability distribution
\begin{equation}
    p(M\SH_{\rm vir}) = \frac{1}{N\SH} \frac{\dd N\SH}{\dd M\SH_{\rm vir}} \;.
\end{equation}
Using the subhalo mass function $\dd N\SH/\dd M_{\rm vir}\SH$ from Ref.~\cite{Hiroshima:2018kfv} and taking the spatial distribution of subhalos to follow an Einasto profile with shape parameter $\gamma=0.854$ and characteristic radius $r_{-2}=245.1\,{\rm kpc}$ (following Ref.~\cite{Ibarra_2019}), we find $n\SH/N\SH = 2.4\times10^{-17}\,{\rm pc^{-3}}$.

For the speed distribution of the subhalos relative to the Solar System, $f\SH(v_\odot\SH)$, we will assume that the subhalos follow the same velocity distribution as DM in the SHM, see Sec.~\ref{sec:signals}. We take the mass-concentration relation, $c\SH(M_{\rm vir}\SH)$, from Ref.~\cite{Moline:2016pbm}. 

\begin{figure}
    \includegraphics[width=\linewidth]{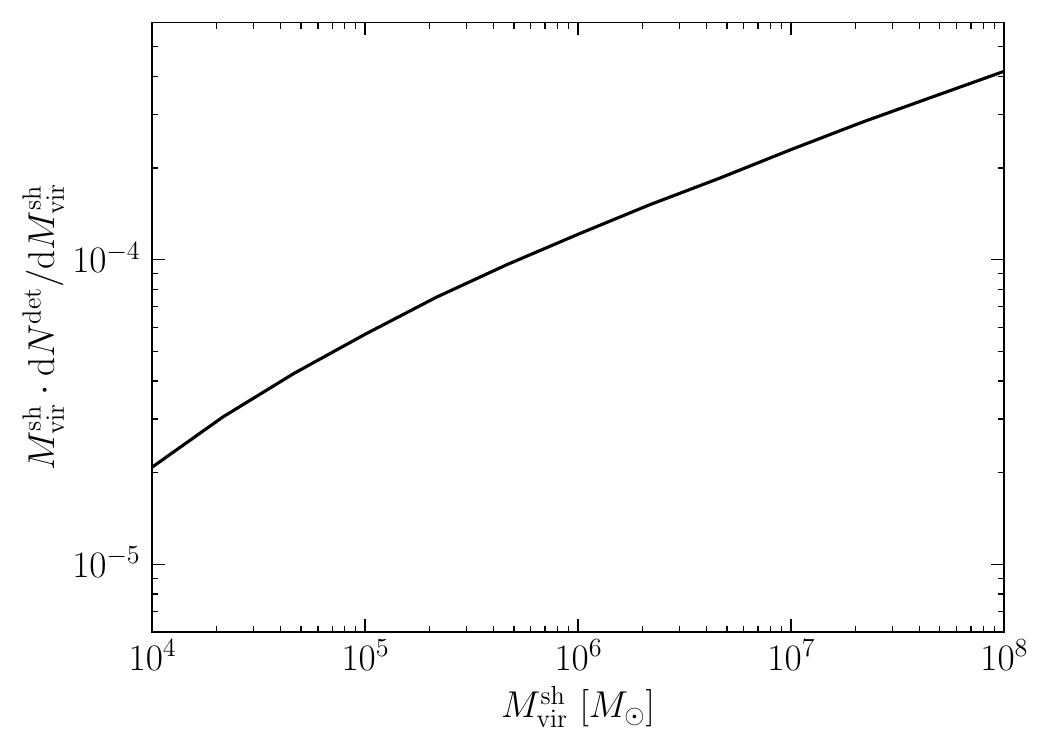}
    \caption{We plot the differential number of detectable subhalos $\dd N^{\rm det}/\dd M_{\rm vir}\SH$, given by Eq.~(\ref{eq:dNdetdM}), as a function of subhalo mass, $M\SH_{\rm vir}$. Here we consider subhalos which could have been detected via a crossing within the past $\Delta T=1\,\text{Gyr}$.}
    \label{Fig:dNdetdM}
\end{figure}

The expected number of detectable subhalo crossings is then
\begin{equation} \label{eq:Ndet}
    N^{\rm det} = \int \frac{\dd N^{\rm det}}{\dd M\SH_{\rm vir}} \dd M\SH_{\rm vir} \;,
\end{equation}
where
\begin{align} \label{eq:dNdetdM}
    \frac{\dd N^{\rm det}}{\dd M\SH_{\rm vir}} &= \frac{\pi n\SH \Delta T}{N\SH} \frac{\dd N\SH}{\dd M\SH _{\rm vir}} \\
    & \times \int f\SH(v_\odot\SH) v_\odot\SH \left[b\SH_{\rm max}(M_{\rm vir}\SH, c\SH(M\SH_{\rm vir}), v_\odot\SH) \right]^2 \dd v_\odot\SH \;, \nonumber
\end{align}
with $b\SH_{\rm max}(M_{\rm vir}\SH, c\SH(M\SH_{\rm vir}), v_\odot\SH)$ the largest impact parameter for given values of $M_{\rm vir}\SH$, $c\SH$, $v_\odot\SH$, $m_\chi\SH$ and $\SIDDSH$ that is discriminable with paleo-detectors. In Fig.~\ref{Fig:dNdetdM}, we plot $\dd N^{\rm det}/\dd M\SH_{\rm vir}$ for our assumptions on the subhalo mass function, mass-concentration relation, subhalo speed distribution, and the spatial distribution of the subhalos in the MW described above under the assumption of our fiducial experimental scenario discussed in Sec.~\ref{sec:Subhaloresults}. Note that such distributions of the subhalos are what is expected for canonical hierarchical structure formation where the Universe was radiation dominated between the end of inflation and redshifts of $z_{\rm eq} \sim 3500$. 

From Fig.~\ref{Fig:dNdetdM} we see that $\dd N^{\rm det}/\dd M\SH_{\rm vir}$ is {\it growing} with $M\SH_{\rm vir}$. In order to estimate $N^{\rm det}$, we must truncate the integral in Eq.~\eqref{eq:Ndet} at a maximum value of $M\SH_{\rm vir}$. MW subhalos with masses $M\SH_{\rm vir} \gtrsim 10^8\,M_\odot$ are expected to contain enough stars to have been detected in astronomical observations; such halos are the so-called dwarf galaxy satellites of the MW. These satellites are catalogued, and their orbits relative to the Solar System can be calculated. Here, we are instead interested in encounters with subhalos that have stellar populations too faint to (presently) be measured in galaxy surveys, hence, we will truncate the integral in Eq.~\eqref{eq:Ndet} at $M\SH_{\rm vir} = 10^8\,M_\odot$. The number of subhalos detectable with a series of paleo detectors with $\max(T^n) \gtrsim \Delta T = 1\,$Gyr is then $N^{\rm det} \sim 10^{-3}$ for our fiducial assumption on the DM mass and scattering cross section, $m_\chi\SH = m_\chi\MW = 500\,$GeV and $\SIDDSH = \SIDDMW = 5 \times 10^{-46}\,{\rm cm}^2$.

The subhalo mass function could be enhanced by orders of magnitude compared to the standard assumption used in the estimate above via a number of mechanisms. For example, a phase transition prior to the era of Big Bang Nucleosynthesis (BBN), a period of early matter domination, or features in the inflationary power spectrum could all lead to large effects in the matter power spectrum (and, in turn, the subhalo mass function) at some range of subhalo masses. Since the chance of finding evidence of a subhalo encounter in a series of paleo-detectors is small in standard cosmology ($\sim 0.1\,\%$ for the assumptions made above), observing evidence for even a single subhalo encounter could offer not only an unprecedented probe of the subhalo mass function at $M_{\rm vir}\SH \ll 10^8\,M_\odot$, but would also provide invaluable hints on pre-BBN cosmology.

\section{Table of Notation}
\label{app:notation_table}
In this Appendix, we collate a comprehensive table of all the symbols used in this paper. We organize this table in the order in which the symbols appear, and include a description for each symbol.

{\renewcommand\arraystretch{1.4}
\begin{longtable*}{C{3cm} L{14cm}}
    \hline \hline
    Symbol & Description\\ 
    \hline
    \multicolumn{2}{c}{Sec.~\ref{sec:intro}}\\
    \hline
    $\varepsilon$ & sample exposure $\varepsilon=M_s\cdot T$\\
    $\Sigma\disk$ & dark disk surface density $\Sigma\disk = {\textstyle \int} \rho_\chi\disk(z)\,\dd z$\\
    $T^{(n)}$ & sample age (optional sample index $n$)\\
    \hline
    \multicolumn{2}{c}{Sec.~\ref{sec:basics}}\\
    \hline
    $x_T$ & damage track length (approximated by the range of a recoiling nucleus in Eq.~\eqref{eq:tracklength})\\
    $E_R$ & energy of recoiling nucleus\\
    $\dd E/\dd x_T$ & stopping power of recoiling nucleus in target material\\
    $\left(\dd R/\dd E_R\right)_{(i)}$ & differential recoil rate per unit target mass, with respect to recoil energy (optional isotope index $i$)\\
    $\dd R/\dd x_T$ & differential recoil rate per unit target mass, with respect to track length\\
    $\xi_i$ & mass fraction of isotope $i$ in target material\\
    $\left(\dd E_R/\dd x_T\right)_{(i)}$ & differential recoil energy, with respect to track length (optional isotope index $i$)\\
    $A$ & isotope mass number \\
    $\sigma_{x_T}$ & track length readout resolution\\
    $M_s^{(n)}$ & sample mass (optional sample index $n$)\\
    $\bm R$ & binned and smeared recoil rate per unit target mass (indexed by bins $i=1,\ldots,N$)\\
    $W$ & window function used for smearing in Eq.~\eqref{eqn:smearing} [defined in Eq.~\eqref{eqn:window}]\\
    $\bm N$ & binned and smeared track length spectrum (indexed by bins $i=1,\ldots,N$)\\
    $\bm n$ & binned and smeared track length spectrum per unit target mass (indexed by bins $i=1,\ldots,N$)\\
    $\mathcal C^{(n)}$ & sample $^{238}{\rm U}$ concentration (optional sample index $n$)\\
    \hline
    \multicolumn{2}{c}{Sec.~\ref{sec:MWhalo}}\\
    \hline
    $m_\chi^{\rm MW/disk/sh}$ & DM particle mass (for Milky Way/dark disk/subhalo)\\
    $m_N$ & mass of target nucleus\\
    $F(E_R)$ & nucleus form factor (taken to be the Helm parametrization in this work)\\
    $\sigma_p^{\rm SI; MW/disk/sh}$ & spin-independent DM-proton cross section (for Milky Way/dark disk/subhalo)\\
    $m_p$ & mass of proton\\
    $\mu_{\chi(N/p)}^{\rm MW/disk/sh}$ & reduced mass of DM-nucleus/proton system $\mu_{\chi(N/p)}=m_\chi m_{N/p}/(m_\chi+m_{N/p})$\\
    $\rho_\chi^{\rm MW/disk/sh}$ & DM energy density (for Milky Way/dark disk/subhalo)\\
    $v_{\rm min}$ & minimum DM speed for given recoil energy $v_\mathrm{min} = \sqrt{m_N E_R/2(\mu_{\chi N}\MW)^2}$\\
    $\eta_\chi^{\rm MW/disk/sh}(v_{\rm min})$ & mean inverse speed, defined in Eq.~\eqref{eqn:recoil_2} (for Milky Way/dark disk/subhalo)\\
    $f^{\rm MW/disk/sh}(\bm v)$ & DM velocity distribution in Solar System frame (for Milky Way/dark disk/subhalo)\\
    $\sigma_v^{\rm MW/disk/sh}$ & DM velocity dispersion (for Milky Way/dark disk/subhalo)\\
    $v_{\rm esc}^{\rm MW}$ & local Galactic escape speed\\
    ${\bm v}_\odot^{\rm MW/disk/sh}$ & velocity of Solar System relative to Milky Way/dark disk/subhalo\\
    ${\bm v}_\oplus$ & orbital velocity of Earth around Sun\\
    \hline
    \multicolumn{2}{c}{Sec.~\ref{sec:darkdisk}}\\
    \hline
    ${\bm v}_v\disk$ & vertical velocity of Solar System relative to dark disk\\
    $Z\disk$ & height of dark disk\\
    $\theta_\odot\disk$ & angle between ${\bm v}_\odot\disk$ and ${\bm v}_v\disk$\\
    $\bar\eta_\chi\disk(v_{\rm min})$ & mean inverse speed averaged over a disk crossing\\
    $\theta_\oplus^\odot$ & angle between ${\bm v}_\odot\disk$ and orbital plane of Earth\\
    ${\bm v}_{\rm rel}\disk$ & velocity of Earth relative to dark disk\\
    $\tilde f\disk$ & DM velocity distribution in rest frame of dark disk\\
    $t_i\disk$ & disk crossing times (indexed by crossings $i$)\\
    \hline
    \multicolumn{2}{c}{ Sec.~\ref{sec:subhalo}}\\
    \hline
    $M_{\rm vir}\SH$ & virial mass of subhalo\\
    $c\SH$ & concentration parameter of subhalo\\
    $z$ & subhalo formation redshift\\
    $\rho_s\SH(M_{\rm vir}\SH, c\SH, z)$ & characteristic density of subhalo, defined in Eq.~\eqref{eq:NFW_rhos}\\
    $r_{\rm vir}\SH(M_{\rm vir}\SH, z)$ & virial radius of subhalo, defined in Eq.~\eqref{eq:NFW_virradius}\\
    $r_s\SH(M_{\rm vir}\SH, z)$ & scale radius of subhalo $r_s\SH=r_{\rm vir}\SH/c\SH$\\
    $\Delta_{\rm vir}(z)$ & critical overdensity required to decouple from cosmic expansion at redshift $z$\\
    $\rho_c(z)$ & critical density at redshift $z$\\
    $b\SH$ & impact parameter of Solar System relative to center of subhalo\\
    $T\SH$ & time of closest approach to center of subhalo\\
    \hline
    \multicolumn{2}{c}{Sec.~\ref{sec:stat}}\\
    \hline
    $\zeta_0$ & parameter controlling signal normalization ($\zeta_0=\SIDDdisk\Sigma\disk$ for dark disk; $\zeta_0=b\SH$ for subhalo)\\
    $\bm\zeta$ & other signal parameters\\
    $\bm\DD$ & track length data set (indexed by bin $i$ and sample $n$)\\
    $\bm\theta$ & nuisance parameters (indexed by $j$)\\
    $\LL$ & log likelihood, defined in Eq.~\eqref{eqn:likelihood} (modified as in Eq.~\eqref{eq:LLGauss} for systematic uncertainties)\\
    $\bar{\bm\theta}$ & central values of nuisance parameters from ancillary measurements (indexed by $j$)\\
    $c_j$ & relative uncertainties of ancillary measurements\\
    $\Phi^{\bm\nu}$ & flux of neutrino backgrounds (for ${\bm \nu} = \{ {\rm solar}~\nu, \, {\rm GSNB}, \, {\rm DSNB} , \, {\rm atm.}~\nu \}$)\\
    ${\bm N}_0(\bm\theta)$ & expected number of tracks from backgrounds and MW halo (indexed by bin $i$ and sample $n$)\\
    $q(\zeta_0)$ & maximum likelihood ratio test statistic\\
    $\hat{\hat{\bm\theta}}$ & maximum likelihood estimator for fixed $\zeta_0$ and $\bm{\zeta}$\\
    $\hat{\bm\theta},\hat\zeta_0$ & maximum likelihood estimators for fixed $\bm{\zeta}$\\
    $\zeta_0^*$ & value of $\zeta_0$ used in Asimov data\\
    $q_{\rm crit}$ & test statistic threshold for discrimination\\
    \hline
    \multicolumn{2}{c}{Sec.~\ref{sec:uncertainty_stats}}\\
    \hline
    $\sigma_i^n$ & uncertainty of the data in the $i$-th bin of the $n$-th sample when applying Eq.~\eqref{eq:LLGauss} \\
    $\sigma_{\rm unc}$ & bin-to-bin modeling uncertainty on background spectra (see Eq.~\eqref{eq:Gausserr})\\
    $\mathcal B_i^n$ & expected number of tracks from backgrounds (indexed by bin $i$ and sample $n$)\\
    \hline
    \multicolumn{2}{c}{Sec.~\ref{sec:Subhaloresults}}\\
    \hline
    $b_{\rm max}\SH(v_\odot\SH)$ & maximum detectable impact parameter for a subhalo crossing\\
    \hline
    \multicolumn{2}{c}{Appendix~\ref{app:subhalo}}\\
    \hline
    $\dd N\SH/\dd M_{\rm vir}\SH$ & subhalo mass function\\
    $n\SH$ & local number density of subhalos\\
    $\Delta T$ & integration time (for calculation in Appendix~\ref{app:subhalo})\\
    $p(M_{\rm vir}\SH)$ & subhalo mass function normalized into probability distribution\\
    $N\SH$ & total number of subhalos in Milky Way\\
    $\gamma$ & shape parameter of Einasto profile\\
    $r_{-2}$ & characteristic radius of Einasto profile\\
    $N^{\rm det}$ & number of detectable subhalo crossings\\
    $\dd N^{\rm det}/\dd M_{\rm vir}\SH$ & differential number of detectable subhalo crossings, with respect to virial mass\\
    \hline\hline
\end{longtable*}
}

\end{appendix}

\bibliography{PaleoBib}

\end{document}